\documentclass[usenatbib,usegraphicx]{aa}
 \usepackage{graphicx}

\usepackage{epstopdf}

\usepackage[usenames]{color}

\newcommand{\dg}{^\circ}

\begin{document}

\title{Theoretical light curves of dipole oscillations in roAp stars}

\author{L. Bigot\inst{1} \and D. W. Kurtz\inst{2}}
\institute{Universit\'e Nice Sophia Antipolis, Observatoire de la C\^ote d'Azur, 
D\'epartement Cassiop\'ee CNRS/UMR 6202, BP 4229, 06304 Nice, France.
\and Jeremiah Horrocks Institute of Astrophysics, University of Central 
Lancashire, Preston PR1\,2HE, UK.}

\date{Received date; accepted date}

\offprints{L. Bigot}

\abstract
{The dipole modes are the most common geometry of oscillations in roAp stars inferred from
photometric measurements and are therefore of special interest for asteroseismic 
purposes.} 
{We present a theoretical and analytical study of the light curves 
associated with dipole ($\ell =1$) pulsations of roAp stars in the framework of 
the revisited oblique pulsator model.} 
{ We describe the light curves in terms of the inclination 
and polarization of the elliptical displacement vector of the dipole modes. We study the influence of the 
magnetic field and rotation on the shape of these light curves for both amplitudes 
and phases.}
 {Despite the inclination of dipole mode with respect to the magnetic axis, we find that the dipole mode  can have 
maxima that are in phase with the magnetic maxima. We apply our formalism to the well-known roAp star HR\,3831 
(HD\,83368) to derive its mode properties. Our results are similar to those 
obtained by time-series spectroscopy. We also consider the cases of three other 
roAp stars, HD\,6532, HD\,99563, and HD\,128898 ($\alpha$\,Cir).
 }
{We demonstrate that the formalism of the revisited oblique pulsator model  is adequate 
to explain the properties of the photometric light curves associated with dipole modes in roAp stars. 
In addition, we show that the coincidence of pulsation and magnetic extrema can also occur for inclined
 modes with respect to the magnetic axis. With the stars considered in this paper, we conclude that the
  polarization of the modes present in roAp stars are quasi linearly polarized.}

\keywords{Stars: magnetic fields -- Stars: oscillations -- Stars: chemically 
peculiar -- Stars: variables: roAp -- Stars: individual: HD\,83368 (HR\,3831), 
HD\,6532, HD\,99563, HD\,128898 ($\alpha$\,Cir).}

\titlerunning{ Theoretical light curves of dipole oscillations  in roAp stars }
\authorrunning{Bigot and Kurtz}
\maketitle

\section{Introduction}

It is well-known that the rapid light variations in Ap stars are caused by trapped 
magneto-acoustic modes. The first detection of these oscillations was made by 
Kurtz (1978, 1982), who clearly showed the presence of high radial order modes 
($n\approx 20-30$) with periods comparable to those of the solar-like stars; these 
periods are now known to range from 6 to 21\,min. The photometric amplitudes of 
the roAp stars, with values of about $0.1-8$\,mmag, are vastly higher than those 
of the solar-like oscillators. It is now common to refer to these objects as roAp 
stars, for {\it rapidly oscillating Ap stars}. To date, over 40 roAp have been 
detected, Kurtz et al. (2006) listing 40 of them.

We can distinguish three fundamental properties of the photometric variability of 
these stars:

\begin{enumerate}

\item The light curves of many roAp stars show a double modulation. The rapid one, 
with a period of a few minutes, arises from the pulsation modes. The long one, 
with a period between days and decades, is the rotation period of the star. To explain 
the presence of the long-term variation, Kurtz (1982) proposed that the pulsation 
mode axis is aligned with the magnetic axis, which is itself inclined to the 
rotation axis so that the observer sees the pulsation modes from an aspect that 
varies with rotation. This is the oblique pulsator model.
\\
\item The second main property of the roAp stars is that the maxima of the 
rotational amplitude modulation of the pulsations occur at the same rotation 
phases as that of the magnetic maxima. This observational finding led Kurtz (1982) to 
justify the inclination of the modes with the rotation axis, simply by considering 
a pulsation axis aligned with the magnetic one. In the 
magnetic Ap stars, it is indeed well-known that the magnetic and rotation axes are inclined with respect to 
each other (e.g. Landstreet \& Mathys 2000); this is the oblique rotator model 
(Stibbs 1950). Kurtz's explanation looked reasonable given the huge, kG-strength 
magnetic fields observed at the surface of these stars, at least two or three 
orders of magnitudes higher than the global magnetic field of the Sun.
\\
\item Finally, the last main property derived from photometry is that for some of 
the roAp stars -- those with geometries such that both magnetic poles come into 
view over the rotation period -- the light curves show a phase jump of $\pm\pi$ 
radians as the pulsation amplitude passes through a minimum. This property is most 
easily explained when the geometrical structure of the mode consists of two 
hemispheres with opposite phases, say a dipole ($\ell=1$) mode, although other, 
higher spherical degree modes may show the same effect for certain viewing 
geometries.

\end{enumerate}

Over the past two decades, many theoretical investigations have studied the 
properties of the roAp stars. The purely geometrical model of Kurtz was improved 
by considering the effects of both the magnetic field and rotation (Dziembowski \& 
Goode 1985; Kurtz \& Shibahashi 1986; Shibahashi \& Takata 1993; Takata \& 
Shibahashi 1995). However, despite these improvements, the magnetic field 
treatment in all these studies was oversimplified. They indeed considered 
a perturbative treatment to solve the magnetic effects in the problem, which is 
unrealistic because in the outer layers of the star the magnetic pressure is higher
than the gas pressure.

In these models, the modes were found to be aligned (or nearly aligned) with the 
magnetic field, in agreement with Kurtz's original proposition of aligned 
pulsation and magnetic axes. The reason behind this alignment is that for high 
overtone pulsations, the magnetic perturbation is much larger than the Coriolis 
one.

On the basis of both photometric observations and theoretical properties, the picture of 
the pulsations in roAp stars over the past two decades has been that of dipole 
oscillations with the pulsation axis aligned with the magnetic axis. The 
inclination of these with respect to the rotation axis thus explains the observed 
long-term modulation of the pulsation amplitude.

Bigot \& Dziembowski (2002) proposed an alternative description of pulsations in roAp stars. 
They found that the axis of pulsation can be inclined by a large angle to the 
magnetic axis. This result comes from a  degenerate perturbative treatment of the rotation 
effects for which they added the quadratic effect of the centrifugal force in the 
problem. This second order effect of rotation was considered negligible in 
previous studies. For high radial order modes, such as those in roAp stars, 
the centrifugal effects can be comparable to the effect of the magnetic field. These 
improvements are crucial for quantitative studies and can change the inclination 
of the mode.

For relatively moderate fields, unlike in the previous studies, there is no 
dominant perturbation acting on the mode, but rather a balance between magnetic 
and rotational perturbations. In their formalism, Bigot \& 
Dziembowski (2002) showed that in the case of dipole modes the pulsation axis lies 
in a plane during the pulsation cycle, with a displacement vector moving with an 
elliptical motion. These dipole modes are fully characterized by three quantities: 
the inclination of the mode plane, the ellipticity of the displacement vector, and 
the frequency. Bigot and Dziembowski applied their model to the roAp star HR\,3831 and based on the 
magnetic measurements (Bagnulo et al. 1999), they showed that, for this star, the 
mode axis is well inclined to both the magnetic and rotation axes.

Considerable progress has since been made in the understanding of pulsation in 
roAp stars by analysing spectroscopic time-series observations. In particular, 
those lead to the possibility of probing the vertical structure of pulsation 
through the atmospheric layers by analysing the pulsation through different lines 
that form at different atmospheric depths. The multiplets observed in spectroscopy 
do not look like those derived by photometry, but have instead stronger non-dipolar 
components. Kochukhov (2006) derived a new geometrical aspect of the mode of 
HR\,3831 using radial velocity measurements and found no significant inclination 
with respect to the magnetic axis. Therefore, he returned to the original 
proposition of Kurtz (1982) and thus questioned the results of Bigot \& 
Dziembowski (2002).

The main objective of this paper is to provide an observational counterpart 
to the model developed in Bigot \& Dziembowski (2002). We derive a formalism to 
calculate analytical light curves associated with an arbitrary combination of dipole 
modes ($m= -1, 0, +1$). The relatively simple geometrical model that we propose is 
particularly useful for analysing photometric pulsation data. We see that it closely 
agrees with both photometric and radial velocity observations. 

In section 2, we give an analytical expression for the light curves of dipole modes in terms of the 
formalism proposed by Bigot \& Dziembowski (2002), i.e. the mode inclination 
and polarization. In section 3, we discuss the correlation between the dipole mode 
inclination and coincidence of pulsation and magnetic extrema. In section 4, we 
consider the effects of a changing magnetic field on these light curves and how it 
affects their amplitudes and phases. Finally, in section 5 we apply these new 
considerations to the cases of four roAp stars HR\,3831, HD\,6532, $\alpha$ Cir, 
and HD\,99563. We see that our formalism agrees well with observations. For 
the particular case of HR\,3831, in the light of a new measurement of the 
inclination angle of the observer, we show that our formalism leads to results 
that are consistent with those of Kochukhov (2006) within the formalism presented 
in the oblique pulsator model of Bigot \& Dziembowski (2002).

\section{The light curve variations of dipole modes}

\subsection{A general expression for non-radial oscillations}

The formalism for light variations in the case of non-radial oscillations is given 
in Dziembowski (1977) (see also Buta \& Smith 1979). We follow the 
formulation of Dziembowski, but with an additional approximation specific to roAp 
stars. In the case of high radial orders, the main contribution of the 
fluctuations to the luminosity $\delta L$ comes from the variations in the 
radiative flux $\delta {\cal F} $. We can reasonably ignore the effects of any changes in the surface element, in terms of both area and orientation. Thus, 
the light variations expressed at the stellar surface may be written as

\begin{equation}\label{eq:deltal}
\frac{\delta L}{L} =\int_0^{2\pi}\int_0^{1}\frac{\delta {\cal
F }}{{\cal F}} h(\mu)\mu d\mu d\phi,
\end{equation}

\noindent where $\mu=\cos\theta$ and $h(\mu)$ is the so-called normalized limb 
darkening function (e.g. Mihalas 1978).

The fluctuation $\delta {\cal F}$ is the unknown quantity that remains to be 
determined. In roAp stars, the modes are strongly influenced by both magnetic and 
inertial forces. The joint effect produced by these two perturbations is non-spherically symmetric, leading to a distortion of  the surface amplitude of the modes. 
Mathematically, this means that $\delta {\cal F }$ is written as a linear 
combination involving spherical harmonics of different degrees $\ell$ and 
corresponding azimuthal orders $m=[-\ell,\ell]$. The eigenmodes are solutions of 
an eigenvalue system. The eigenvalues are the frequencies, and the eigenvectors 
are the coefficients of the spherical harmonic expansion.

In this context, Kurtz (1992) proposed that the light variations of roAp 
stars could be interpreted in terms of axisymmetric modes ($m=0$) about the magnetic field, whose 
angular dependence is described by a linear combination of $\ell$ ($=0,1,2,3$). 
This combination of spherical harmonics was used to take into account the 
distortion produced by the magnetic field. Dziembowski \& Goode (1996) and Bigot 
et al. (2000) showed in the case of dipole magnetic fields that the distortion of 
the eigenvectors is quite weak as long as the field strength remains $< 1$\,kG and 
can be regarded as a small correction to the problem. Cunha \& Gough (2000), Saio 
\& Gautschy (2004), Saio (2005), and Cunha (2006) explored the effects of stronger 
fields and showed that the surface amplitudes of the modes are strongly modified by 
kilogauss field strengths as found in roAp stars. However, photometric observations show dominant triplets, which mean that either the magnetic distortion is small or the geometric cancellation (integration over the stellar disk) of components $\ell>1$ leads the observer to see the distorted mode as a dipole. In the following, we assume a negligible magnetic distortion in order to develop an analytical formalism, keeping in mind that it might not be applicable to all roAp stars.


In this respect, the perturbation of the radiative flux may be written as

\begin{equation}\label{eq:deltaf}
\frac{\delta {\cal F}}{{\cal F}} (\mu_R,\phi_R,t) =\epsilon {\it
f}_{\ell}\left [\sum_{m=-\ell}^{\ell} a_m
Y_{\ell}^m(\mu_R,\phi_R)\right ] e^{{\rm i }\omega t},
\end{equation}

\noindent where $a_m$ are the coefficients of the coupling and $\omega$ the frequency. 
These quantities are, respectively, the eigenvectors and eigenvalues of an 
homogenous system (e.g. Dziembowski \& Goode 1985; Kurtz \& Shibahashi 1986; Bigot 
\& Dziembowski 2002). They depend on the magnetic and 
rotation strengths and orientations. The quantity ${\it f}_{\ell}$ is a function 
that characterizes the radial dependence of $\delta {\cal F}$ and is expressed 
here at the stellar surface. The coefficient $\epsilon$ is unknown in the linear 
treatment of the oscillations, but is a small parameter so the 
perturbation formalism is valid. The $Y_{\ell}^m$ are the usual spherical 
harmonics expressed in the rotation reference system, i.e. $\mu_R=1$ corresponds 
to the rotation axis.

We emphasize that the present approach differs from that of Kurtz (1992) since we 
consider that the distortion of the mode is described by a linear combination, not 
for $\ell$, but for the azimuthal order $m$. We stress that because of the 
inclination between the magnetic and rotational perturbations the coupling of 
modes for different azimuthal orders $m$ can be strong and cannot be ignored in 
roAp stars.

To proceed further and compare with observations, we have to write the flux 
variations described by Eq.~(\ref{eq:deltaf}) in the inertial reference system whose polar axis coincides 
with the line-of-sight of the observer. This operation is done via the well-known 
relation for spherical harmonics

\begin{equation}\label{eq:trans}
Y_{\ell}^m(\mu_R,\phi_R) =\sum_{j=-\ell}^{\ell} d_{mj}^{(\ell)}(i)
 Y_{\ell}^j(\mu_L,\phi_L),
\end{equation}

\noindent where $(\mu_L,\phi_L)$ are the coordinates in the new (observer's) reference 
system, $i$ the angle between the rotation axis and the line-of-sight, and 
$d_{mj}^{(\ell)}$ coefficients that are expressed in terms of the Jacobi 
polynomials (e.g. Edmonds 1960). Moreover, the observer being in an inertial 
frame, we have to transform the coordinate system as $\phi_R\rightarrow\phi_R -
\Omega t$, where $\Omega$ is the rotation frequency of the star. Therefore, 
after substituting  Eq.\,\ref{eq:trans} into Eq.\,\ref{eq:deltaf}, and integrating over the stellar disk (only $j=0$ remains in Eq.\,\ref{eq:trans}) the expression for the 
fluctuation in luminosity is given by

\begin{equation}\label{eq:deltalbis}
\frac{\delta L}{ L} = 2\pi\epsilon {\it f}_{\ell}\left (
\frac{2\ell+1}{4\pi}\right )^{1/2} b_{\ell}\sum_{m=-\ell}^{\ell}
A_{\ell}^m\cos (\omega -m\Omega )t,
\end{equation}

\noindent where

\begin{equation}
A_{\ell}^m =\left [\frac{(\ell-m)!}{(\ell+m)!}\right ]^{1/2} P_{\ell}^m
(\cos i)\,\,a_m,
\end{equation}

\noindent and

\begin{equation} b_{\ell} =
\int_0^1 h(\mu_L) P_{\ell}(\mu_L)\mu_L d\mu_L.
\end{equation}

Since we work at a given value of $\ell$ and consider only relative 
fluctuations in the following sections, we are not interested in the value of the 
constant in front of the sum in Eq.\,\ref{eq:deltalbis}, and just consider

\begin{equation}\label{eq:deltalbisbis}
\frac{\delta L}{L}\propto\sum_{m=-\ell}^{\ell} A_{\ell}^m\cos
(\omega -m\Omega )t.
\end{equation}

\noindent This expression clearly shows the double modulation observed in roAp stars, a 
rapid one of period $2\pi/\omega$ due to the oscillation and a long one of period $2\pi/\Omega$ due to the rotation of the star. In the 
frequency domain, it leads to a $(2\ell+1)$ multiplet of amplitudes $A_m$ equally 
spaced by the frequency of rotation.

\subsection{Light curve variations of dipole modes}

The dipole modes have the most common geometry observed in photometry in roAp stars.  They are therefore of special interest to asteroseismic studies. 
In the rest of this work, we will consider the fluctuations of luminosity in Eq. \ref{eq:deltalbisbis}  for generalized dipole ($\ell=1$) modes, following the formalism developed in Bigot \& Dziembowski (2002). We will present an analytical formulation of the light curves for both amplitudes and phases.

\subsubsection{Properties of dipole modes}

As shown in Bigot \& Dziembowski (2002), the dipole mode axis lies in a fixed 
plane during the pulsation cycle with an elliptical
displacement vector.
\begin{figure*}[t!]
\begin{center}
\includegraphics[width=12cm,angle=0]{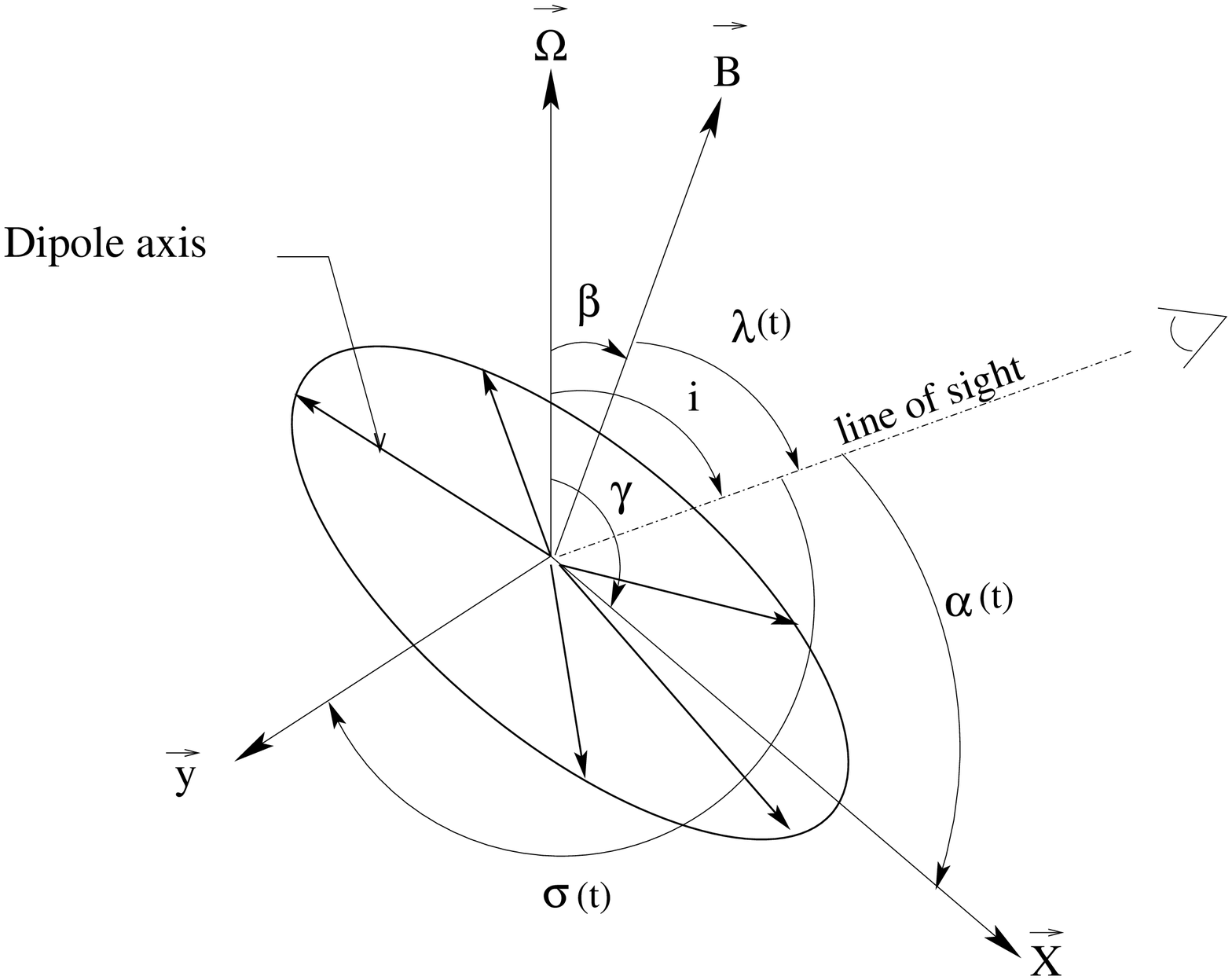}
\caption{The oblique pulsator model for dipole modes (Bigot \& Dziembowski 2002). 
During the pulsation cycle, the axis of the dipole mode lies in a fixed plane 
that is inclined to both the rotation and magnetic axes. The shape of the 
displacement vector is elliptical. The two axes of the ellipse are the $X-$axis in 
the $\bf (B,\Omega)$ plane and the perpendicular $y-$axis. The x-axis lies in the rotation equatorial plane.The line-of-sight of 
the observer makes an angle $\alpha (t)$ with the former, and $\sigma(t)$ with the 
latter. The magnetic axis is inclined by the angle $\lambda(t)$ to the line-of-sight.} \label{fig:dipole}
\end{center}
\end{figure*}
The orientation and shape of the ellipse are defined by its two axes: the $X$-axis 
in the plane defined by the vectors $({\bf B},{\bf\Omega})$ and the perpendicular 
$y-$axis. The inclination of the $X$-axis of the ellipse with respect to the 
rotation axis is given by

\begin{equation}
\gamma =\arctan\left [\frac{a_{+1}-a_{-1}}{\sqrt{2}a_0}\right ].
\end{equation}

\noindent The elliptical displacement of the mode is characterized by the 
polarization angle

\begin{equation}
\psi=\arctan\left [\frac{a_{+1}+a_{-1}}{a_{+1}-a_{-1}}\sin\gamma\right ],
\end{equation}

\noindent which measures the ratio of the two axes of the ellipse. The 
projected amplitudes of the displacement vector onto the X- and y-axes are then 
$\cos\psi$ and $\sin\psi$, respectively. We note that $\psi=0$ corresponds to a 
mode that is linearly polarized in the plane $({\bf B},{\bf\Omega})$, whereas 
$\psi=\pm\pi/2$ corresponds to a mode that is linearly polarized along the y-axis.  \\

 The special cases are:
\begin{itemize}

\item ($|\gamma|=0,\psi=0$) for a mode linearly polarized along the rotation axis, 
i.e. an $m_R=0$ mode in the rotation system;

\item ($|\gamma|=\pi/2,\psi=\pm\pi/4$) for a mode circularly polarized in a plane 
orthogonal to the rotation axis, i.e. an $m_R=\pm 1$ mode in the rotation system;

\item ($|\gamma|=\beta,\psi=0$) for a mode linearly polarized along the magnetic 
axis, i.e. an $m_B=0$ mode in the magnetic system;

\item ($|\gamma|=\pi/2-\beta,\psi=\pm\pi/4$) for a mode circularly polarized in a 
plane orthogonal to the magnetic axis, i.e. an $m_B=\pm 1$ mode in the magnetic system.

\end{itemize}

\subsubsection{Amplitudes and phases of the light curves}

The light curve associated with a dipole mode may be obtained from 
Eq.\,\ref{eq:deltalbisbis} with $\ell=1$. We write it in a form similar to that in 
Kurtz (1992)

\begin{equation}\label{eq:lighttab}
\frac{\delta L}{L}\propto A(t)\cos\omega t + B(t)\sin\omega t,
\end{equation}

\noindent but with different coefficients $A(t)$ and $B(t)$,
\begin{equation}
A(t) = a_0\cos i +\frac{a_{+1}-a_{-1}}{\sqrt{2}}\sin i\cos\Omega t,
\end{equation}

\noindent and

\begin{equation}
B(t) =\frac{a_{+1}+a_{-1}}{\sqrt{2}}\sin i\sin\Omega t.
\end{equation}

\noindent These expressions can be written in terms of $\gamma$ and $\psi$ as

\begin{equation}
A(t) =\cos\psi\left (\cos\gamma\cos i +\sin\gamma\sin i\cos\Omega t\right ),
\end{equation}

\noindent and

\begin{equation}
B(t) =\sin\psi\sin i\sin\Omega t,
\end{equation}

\noindent where we have normalized each coefficient $a_m$ (which are solutions  for a homogeneous system)  by $\sqrt{a_0^2+a_1^2+a_{-1}^2}$. 

From the observer's point of view, the  relevant quantities are the orientations of the 
$X-$ and $y-$axes with respect to the line-of-sight, rather than the orientation 
with respect to the rotation axis (z-axis).  It is then more natural to consider the angles 
made by these two axes to the line-of-sight instead of the angles $\gamma$ and 
$i$.

To define these angles, we consider the three directions of the problem that are 
defined by the three following unit vectors:

\begin{enumerate}

\item the direction of the inertial observer

\begin{equation}
{\bf n_{obs}} = (\sin i\cos\Phi (t),\sin i\sin\Phi (t),\cos i),
\end{equation}

\noindent where $\Phi(t) =\Omega t$ is the rotational phase;
\\
\item the direction of the $X$-axis

\begin{equation}\label{eq:nx}
{\bf n_X} = (\sin\gamma,0,\cos\gamma);
\end{equation}

\item the direction of the $y-$axis

\begin{equation}\label{eq:nx}
{\bf n_y} = (0,1,0).
\end{equation}
\end{enumerate}

\noindent The angle $\alpha (t)$ between the $X$-axis of the ellipse and the line-of-sight 
is then defined by

\begin{equation}
\cos\alpha (t) = {\bf n_X}.{\bf n_{obs}} =\cos i\cos\gamma +\sin\gamma
\sin i\cos\Phi (t),
\end{equation}

\noindent and the angle $\sigma (t)$ between the $y$-axis of the ellipse and the 
line-of-sight is

\begin{equation}
\cos\sigma (t) = {\bf n_y}.{\bf n_{obs}} =\sin i\sin\Phi (t).
\end{equation}

\noindent The expressions for the coefficients $A(t)$ and $B(t)$ reduce to simple forms

\begin{eqnarray}
& & A(\Phi) =\cos\psi\cos\alpha (\Phi)\\
& & B(\Phi) =\sin\psi\cos\sigma (\Phi).
\end{eqnarray}

\noindent These quantities then have a simple physical explanation: they correspond to the 
lengths of the $X-$ and $y-$ axes of the ellipse ($\cos\psi$ and $\sin\psi$, 
respectively, in the mode plane) projected onto the line-of-sight. The amplitude 
and phase of the signal related to the dipole modes are finally only functions of three 
quantities: the eccentricity of the ellipse ($\psi$) and its orientation with 
respect to the observer ($\alpha$, $\sigma$).

Following Kurtz (1992), we write the light curve in terms of its amplitude $R(t)$ 
and phase $\Psi (t)$

\begin{equation}\label{eq:deltal}
\frac{\delta L}{L} (t)\propto R (t)\cos (\omega t -\Psi (t)),
\end{equation}

\noindent with

\begin{equation}\label{eq:amp}
R(t) =\sqrt{ A^2(t) + B^2(t)}
\end{equation}

\noindent and

\begin{equation}\label{eq:pha}
\Psi(t) = \arctan\left [\frac{B(t)}{A(t)}\right ].
\end{equation}

\begin{figure*}[t!]
\begin{center}
\includegraphics[width=8.5cm,angle=0]{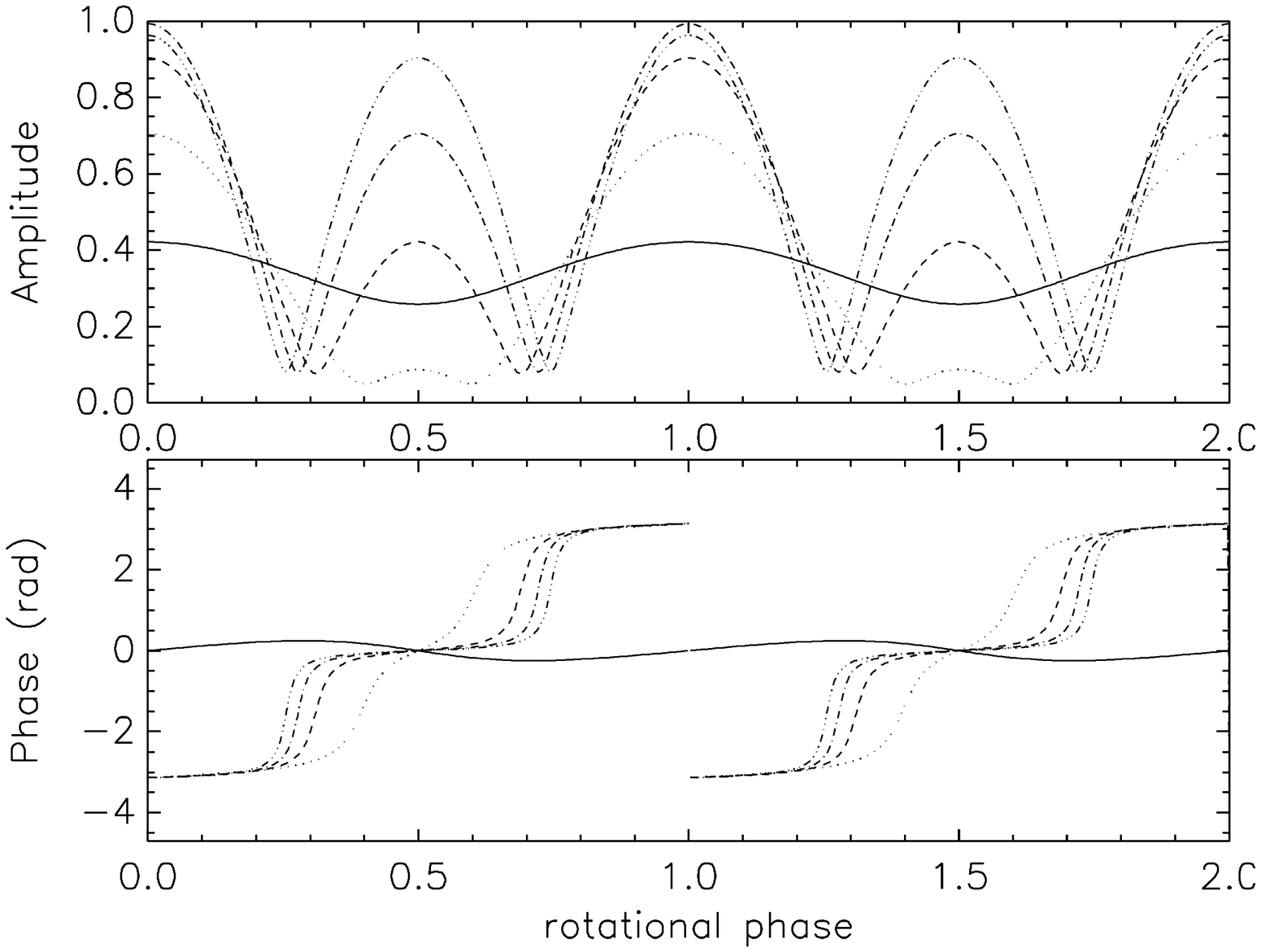}
\includegraphics[width=8.5cm,angle=0]{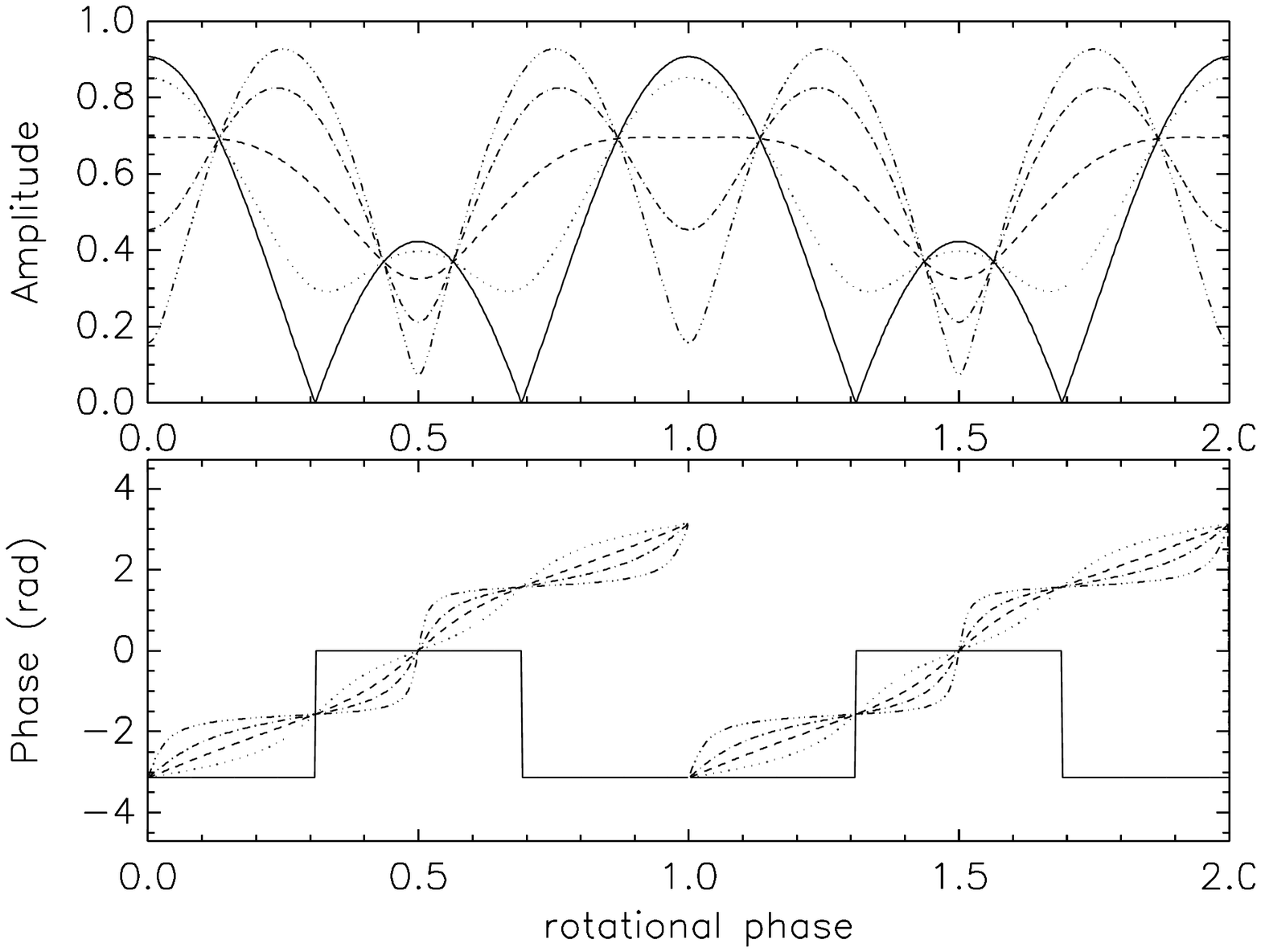}
\caption{Shapes of the light curves influenced by different 
inclinations and polarizations of the dipole mode.  In both panels, the inclination of the observer is $i=70^\circ$. (Left panel) We consider a given polarization of the mode $\psi=5$$^\circ$, and compute 
curves for various inclinations of the mode plane: $\gamma =$ 5$^\circ$ (---), 
25$^\circ$ ($\cdots$), 45$^\circ$ ($---$), 65$^\circ$ ($-\cdot-$), 85$^\circ$ ($-
\cdots-$). It is clear that for all inclinations but one ($\gamma=5^\circ$), the 
phase jumps $\pi$ radians exactly at the amplitude minima since the modes show the 
two hemispheres during the rotation period (i.e., this verifies the conditions given in Eq.~\ref{eq:condition}). The position of the minima at $\Phi_m$ are shifted by a 
change in the inclination of the mode, as one can check in Eq.\,\ref{eq:extrema}. 
(Right panel) Shapes of the light curves influenced by different 
polarizations of the dipole mode. We consider a given inclination of the mode $\gamma=45$$^\circ$, and 
compute curves for various polarizations $\psi$ of the dipole axis: 
$\psi=0$$^\circ$ (---), $20$$^\circ$ ($\cdots$), $40$$^\circ$ ($--- $), 60$^\circ$ 
($-\cdot-$), 80$^\circ$ ($-\cdots-$). In both panels, we have 
represented two periods of rotation. } 
\label{fig:delta}
\end{center}
\end{figure*}

\noindent The behaviour of the amplitude $R$ and phase $\Psi$ for different mode 
inclinations and polarizations is represented in Fig.\,\ref{fig:delta}.

\section{Extrema of the rotationally-modulated amplitudes of dipole modes and their coincidence 
with magnetic maxima}

The number and nature of the extrema of the rotationally-modulated amplitudes 
depend on the aspect of the dipole mode seen by the observer. In general, within 
one period of rotation, $\Phi=[0,2\pi]$, we count four extrema located at

\begin{equation}\label{eq:extrema}
\Phi =0\,[\pi] \hspace{.4cm}{\rm and} \hspace{.4cm}\Phi\equiv
\Phi_o^{\pm} =\pm\arccos\left (\frac{\xi}{\tau^2 -1}\right ),
\end{equation}

\noindent where $\tau =\tan\psi/\sin\gamma$ and $\xi =\cot\gamma\cot i$. The 
presence of four extrema of the envelope, despite the dipolar nature of the mode, 
is a direct consequence of the definition of $R(\Phi)$ introduced in 
Eq.\,\ref{eq:amp}.

In some special cases (polarization and/or inclination), the observer might see 
only two extrema. This may occur for example if $|\xi|\geq |\tau^2 -1 |$. In that 
case, the observer sees extrema only at $\Phi = 0 [\pi]$. The inclination of the 
observer relatively to the mode axis is such that the same hemisphere is always 
seen as the star rotates. An illustration is given in the appendix in the special 
case of $\psi=0$ modes.

Another case for the presence of only two extrema is found when the displacement vector 
of the mode is always perpendicular to the ${\bf (B,\Omega)}$ plane, i.e. when 
$|\psi|=\pi/2$. In that case, there are no extrema at $\Phi = 0\,[\pi]$, but only 
at $\Phi=\Phi_o =\pm\pi/2$.

In the following, we focus our discussion of the extrema located at $\Phi= 0 
[\pi]$. The objective is to determine under which conditions the pulsations show 
maxima in phase with the magnetic maxima.

\subsection{The extrema of the magnetic field}

Owing to its inclination with respect to the rotation axis, the magnetic field 
is also modulated as the star rotates. The modulation depends on the orientation 
of its axis with respect to the line-of-sight of the observer. This describes the 
standard, well-known oblique rotator model (Stibbs 1950). Here we present this 
model in our notation.

The direction of the magnetic field axis with the rotation axis is defined by the 
unit vector

\begin{equation}
{\bf n_{B}} = (\sin\beta,0,\cos\beta),
\end{equation}

\noindent where $\beta$ is the angle between the two axes. From the observer's 
line-of-sight, the magnetic axis is inclined by an angle $\lambda$ defined by

\begin{equation} \cos\lambda (\Phi) = {\bf n_{B}}.{\bf n_{obs}} =
\cos\beta\cos i +\sin\beta\sin i\cos\Phi (t).
\end{equation}

\noindent Hence, the magnetic modulation due to the rotation of the star is

\begin{equation}
\langle B\rangle\propto\cos\lambda (\Phi).
\end{equation}

\noindent This magnetic variation is a sinusoid with two maxima at $\Phi=0$ and $\Phi=\pi$. 
These maxima can be positive or negative depending on the value of $\lambda$ at 
$\Phi=0$ or $\Phi=\pi$.

The condition that has to be met for a positive maximum at $\Phi=0$ and a negative minimum at $\Phi=\pi$ is

\begin{equation}\label{eq:mag1}
|\lambda (0)|\leq \frac{\pi}{2} \leq |\lambda (\pi)|\hspace{0.1cm}
{\rm or\, equivalently}\hspace{0.1cm} |\beta -i|\leq \frac{\pi}{2} \leq
|\beta+i|.
\end{equation}

\noindent The first inequality ensures that there is a positive maximum at 
$\Phi=0$, i.e. $|\lambda(0)|\leq \frac{\pi}{2}$, and the second one ensures that 
there is a second negative maximum, i.e. $|\lambda(\pi)|\geq \frac{\pi}{2}$.

\subsection{The extrema of pulsation at $\Phi=0\,[\pi]$}

We now consider the nature of the pulsation extrema at the same rotation phases as 
those of the magnetic extrema. The first requirement to have magnetic and 
pulsation maxima in phase is obviously that the rotational envelope shows extrema 
at $\Phi = 0\, [\pi]$. This means that the mode must have a non-zero amplitude in 
the magnetic plane $({\bf B,\Omega})$, which requires that $|\psi|\neq\pi/2\, 
[\pi]$. If this condition is satisfied, the dipole mode always has two extrema in 
phase with the magnetic extrema at $\Phi = 0\, [\pi]$, since the $X-$axis of the 
ellipse is within the magnetic plane. The question is then to know whether these 
extrema are maxima or minima. The answer depends on both the polarization $\psi$ 
of the dipole mode and its orientation toward the observer.

At the rotation phase $\Phi = 0\,[\pi]$, the angles between the $X$-axis of the 
ellipse and the line-of-sight are

\begin{equation}
\alpha (0) =\gamma - i \hspace{.4cm} {\rm and} \hspace{.4cm}
\alpha (\pi) =\gamma + i.
\end{equation}

\noindent The $y$-axis of the ellipse is perpendicular to the line-of-sight 
($\sigma =\pm\pi/2$) so its projected component becomes zero. The condition under which  $R(\Phi)$ 
shows a maximum at $\Phi = 0 $ is easily derived by a simple analysis of 
Eq.\,\ref{eq:amp}. We find that

\begin{equation}\label{eq:max0}
\left (\tan\psi\right )^2 <\frac{\cos\alpha (0)(\cos\alpha (0) -
\cos\alpha(\pi))}{2 \cos^2\sigma(\pi/2)}.
\end{equation}

\noindent A similar expression is obtained at the phase $\Phi=\pi$,

\begin{equation}\label{eq:maxpi}
\left (\tan\psi\right )^2 <\frac{\cos\alpha (\pi) (\cos\alpha (\pi) -
\cos \alpha(0))}{2 \cos^2\sigma(\pi/2)}.
\end{equation}

\noindent The conditions Eqs\,\ref{eq:max0} and \ref{eq:maxpi} are restrictive, in the sense 
that for a given position of the observer relative to the mode axis (i.e. X-axis 
at $\Phi=0\, [\pi$]), there is upper limit to the strength of the polarization allowed. In 
other words, the displacement vector has to stay close to the ${\bf (B,\Omega)}$ 
plane in order to show a maximum in phase with the magnetic one.

\begin{figure*}
\begin{center}
\includegraphics[width=17cm,angle=0]{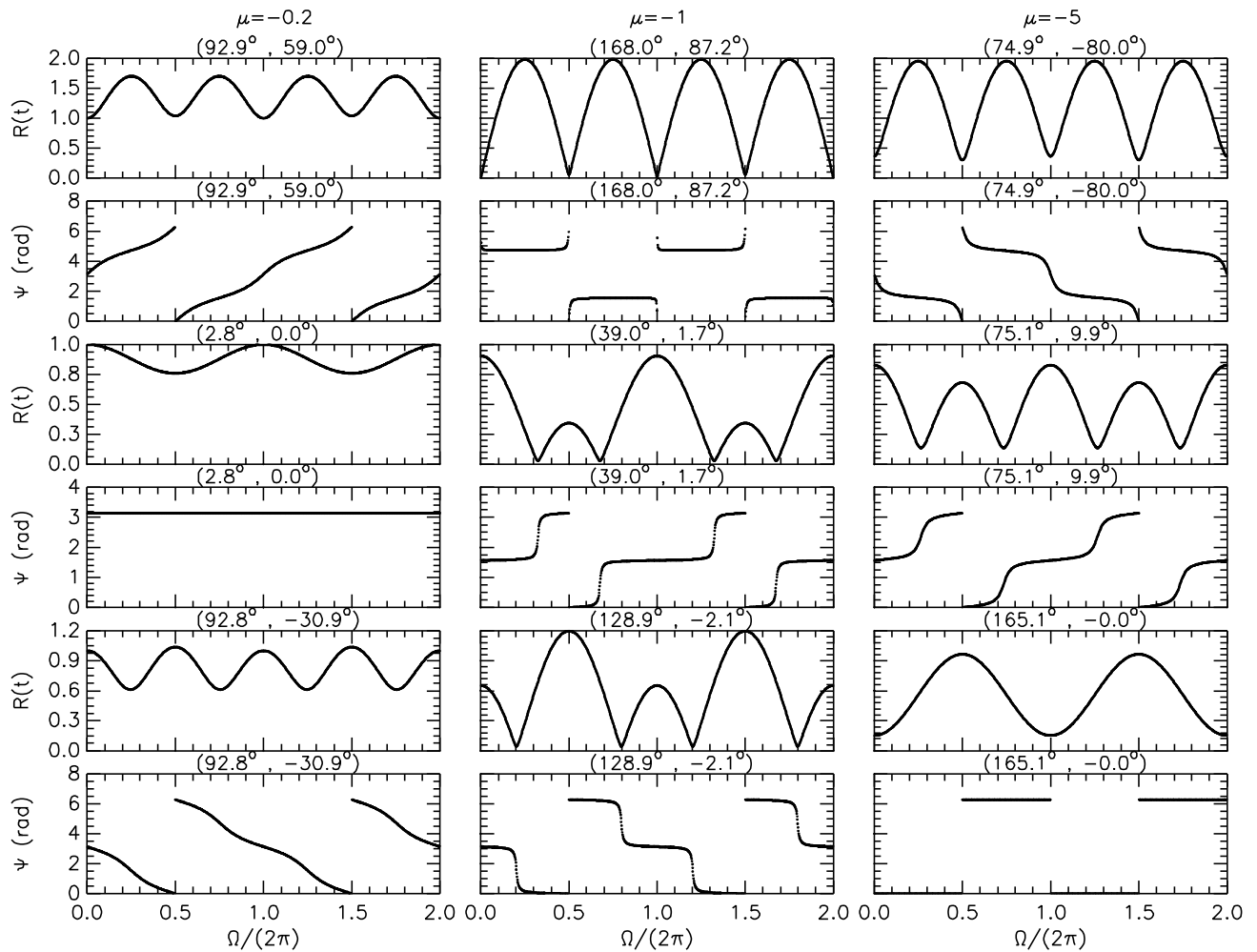}
\caption{Representation of the rotational variation of the light curves in 
amplitude and phase for three dipole modes on the six rows, and for three regimes 
of magnetic field in the three columns, $\mu=-0.2,\,-1.0\,$, and $-5.0$, where 
$\mu$ measures the ratio of the magnetic to centrifugal shifts of the pulsation 
frequencies. The amplitudes are normalized so that the maximum is equal to unity. 
In all cases, we choose $i=70$$^\circ$ and $\beta=12$$^\circ$. The ratio of the Coriolis to centrifugal shifts of frequencies is $\chi=0.01$. See the text 
for a description of the properties of the three dipole modes. The values (in degrees) of the inclination  and polarization are given for  each mode in the brackets $(\gamma,\psi)$ above the panels. }
\label{fig:modul10}
\end{center}
\end{figure*}

It is important to note that these relations put some constraints on the 
polarization of the dipole mode. However, they allow modes that are (well) inclined to the 
magnetic field and have maxima in phase with the magnetic ones. Until now, it has 
been common to assume that the coincidence between the pulsation and magnetic 
maxima implies that modes are linearly polarized along the magnetic axis, say $\psi=0$ and 
$\gamma =\beta$. Even though this case does lead to such a coincidence, it is 
certainly not the only possibility. The relations in Eqs \ref{eq:max0} and 
\ref{eq:maxpi} indeed do not exclude modes with circular polarizations, $\psi=\pm\pi/4$, 
and even modes that are more elongated along the $y$-axis ($\pi/4\leq |\psi| 
<\pi/2$).

This result is not obvious: modes that have the largest amplitude of their 
displacement vector in a direction perpendicular to the magnetic plane ${\bf 
(B,\Omega)}$, i.e. modes such as ($|\psi|>\pi/4$), can show some maxima in phase 
with the magnetic maxima. This apparent contradiction can be easily explained because 
the observer does not see the real amplitude of the displacement vector but rather 
the projection of the $X$- and $y$-axes of the ellipse onto the line-of-sight. 
What really matters is the amplitude of the {\it projection} of the X-axis 
relative to the amplitude of the projection of the $y$-axis. The mode shows a 
maximum in phase with the magnetic maximum if the projection of its displacement 
in the plane (${\bf B,\Omega}$) is larger than the one perpendicular to this 
plane.

To illustrate this, we consider an example: a mode with $\psi=55$$^\circ$, $\gamma=50$$^\circ$, and 
$i=10$$^\circ$. This mode is more elongated along the $y$-axis than the $X$-axis, 
since $\sin\psi>\cos\psi$. The inequalities in equations Eqs \ref{eq:max0} and 
\ref{eq:maxpi} are fulfilled, and the two extrema at $\Phi=0$ and $\Phi=\pi$ are 
maxima. The projection of the $X$-axis onto the line-of-sight is $a(\Phi=0)=0.43$, 
whereas the projection of the $y$-axis is $b(\Phi=\pi/2) = 0.14$. Hence, the mode 
shows a maximum at $\Phi=0$ because the projection of the displacement vector 
component in the plane {\bf $(\bf B,\Omega)$} is larger than its component 
perpendicular to this plane ($y\,-$ direction).

The two extrema have unequal amplitudes because the two poles of the mode have 
different inclinations ($\alpha(0)\neq\alpha(\pi)$) with respect to the line-of-
sight. The ratio of their amplitudes is given by

\begin{equation}\label{eq:ratio}
\frac{R(\pi)}{R(0)} =\left |\frac{\cos\alpha(\pi)}{\cos\alpha (0)}\right | = \left 
| \frac{\cos(i+\gamma)}{\cos(i-\gamma)}\right |.
\end{equation}

\noindent This ratio is independent of $\psi$ since at $\Phi=0\,[\pi]$ the small 
axis of the ellipse is perpendicular to the line-of-sight (null projection). We 
note that for some special values of the inclination of the observer, 
$i=0\,[\pi/4]$, this ratio equals one, regardless of the inclination of the mode. 
These cases are unfortunate for asteroseismology because they cannot provide 
information on the mode.

\section{Magnetic versus rotation effects on the light curves}

We now consider the influence of a varying magnetic field and rotation on the 
aspect of the light curves. At a given rotation frequency $\Omega$, a change in 
the magnetic field strength modifies both the inclination and polarization of the mode, 
which thereby changes its aspect seen by the observer. The results on the combined 
effects of a magnetic field and rotation on the dipole mode property are discussed 
in detail in Bigot \& Dziembowski (2002). To relate the present work to 
these results, we keep the same values for the mode parameters and the same 
notation as Bigot \& Dziembowski (2002). The magnetic field strength is 
parameterized by the value of $\mu$, which measures the ratio of the magnetic 
to centrifugal shifts in frequencies. We emphasize that no assumption is made 
regarding the method used to solve the magnetic shift of frequencies that enters 
into the $\mu$ parameter. The values of $|\mu|> 1$ correspond to a predominant 
magnetic field regime, and $|\mu| < 1$ corresponds to a regime where the centrifugal 
force dominates. In Fig.\,\ref{fig:modul10}, we use $\mu = -0.2$, $-1.0$ and $-
5.0$. The ratio of Coriolis to centrifugal shifts of frequencies is kept to a 
constant value of $0.01$, which is close to the real value corresponding to the 
roAp star HR\,3831.

For dipole modes, the eigenvalue system leads to three orthogonal solutions that 
are represented on each row of Fig.\,\ref{fig:modul10}. The aspect of these three 
modes is strongly modified when the magnetic field strength increases, as one can 
see from left to right :

\begin{enumerate}

\item In the low magnetic field regime ($|\mu| < 1$, left column), each eigenmode 
is almost described by a single spherical harmonic in the rotation system ($m_R= 
0,\pm 1$). The mode that is almost an $m_R=0$ is not modulated because its axis 
nearly coincides with the rotation axis. The phase of the modes that are almost 
$m_R =\pm 1$ modes show this characteristic straight line variation for the phase.

\item The most complicated situation arises when magnetic and rotation effects are 
comparable ($|\mu|=1$, middle column), where the coupling between $m$-components of 
the mode is the strongest. In that case, it is impossible to identify a mode with a 
single value of $m$.

\item In the case of strong magnetic field ($|\mu| > 1$, right column), the modes 
are again almost described by a single value of $m$ but now in the magnetic 
reference system. The one that is almost described by an $m_B=0$ mode shows a 
constant phase except for phase reversals at amplitude minima. This jump 
corresponds to the moment when the node of the oscillations passes through the 
plane $({\bf i,\Omega})$. The two other orthogonal modes are almost circularly 
polarized $(\psi=\pm\pi/4)$ and tend to show characteristic monotonic linearly 
increasing or decreasing phases.

\end{enumerate}

For a given mode (a given row in Fig.\,\ref{fig:modul10}), the light curve aspect 
changes drastically with the value of $\mu$. This is because the mode tends to be 
aligned with ($m_B=0$), or orthogonal ($|m_B|=1$) to, the magnetic axis as the 
field strength increases. This changes both the numbers and the amplitudes of the 
maxima of the rotational envelope, as well as the polarization. The mode that is 
almost an $|m_R|=1$ mode in the rotation frame for $\mu <1$ becomes an $m_B=0$ 
mode in the magnetic frame when $\mu >1$. Its phase curve changes from nearly 
straight lines to constant phase with jumps.

We emphasize that there is a saturation effect once the magnetic field 
(equivalently $\mu$) reaches a certain value ($\sim 4-5$). Once the field is large 
enough, the three eigenmodes are either aligned or perpendicular to the magnetic 
axis and an increase in the field strength no longer has an effect on the aspect 
of the mode seen by the observer. Therefore, it becomes impossible to constrain 
the field strength above a certain limit by fitting the relative mode amplitudes. 
A tighter constraint is provided by the frequency spacing.

\section{Applications to four roAp stars}

\subsection{Application to HR\,3831}

\begin{figure*}[t!]
\begin{center}
\includegraphics[width=6.5cm,angle=0]{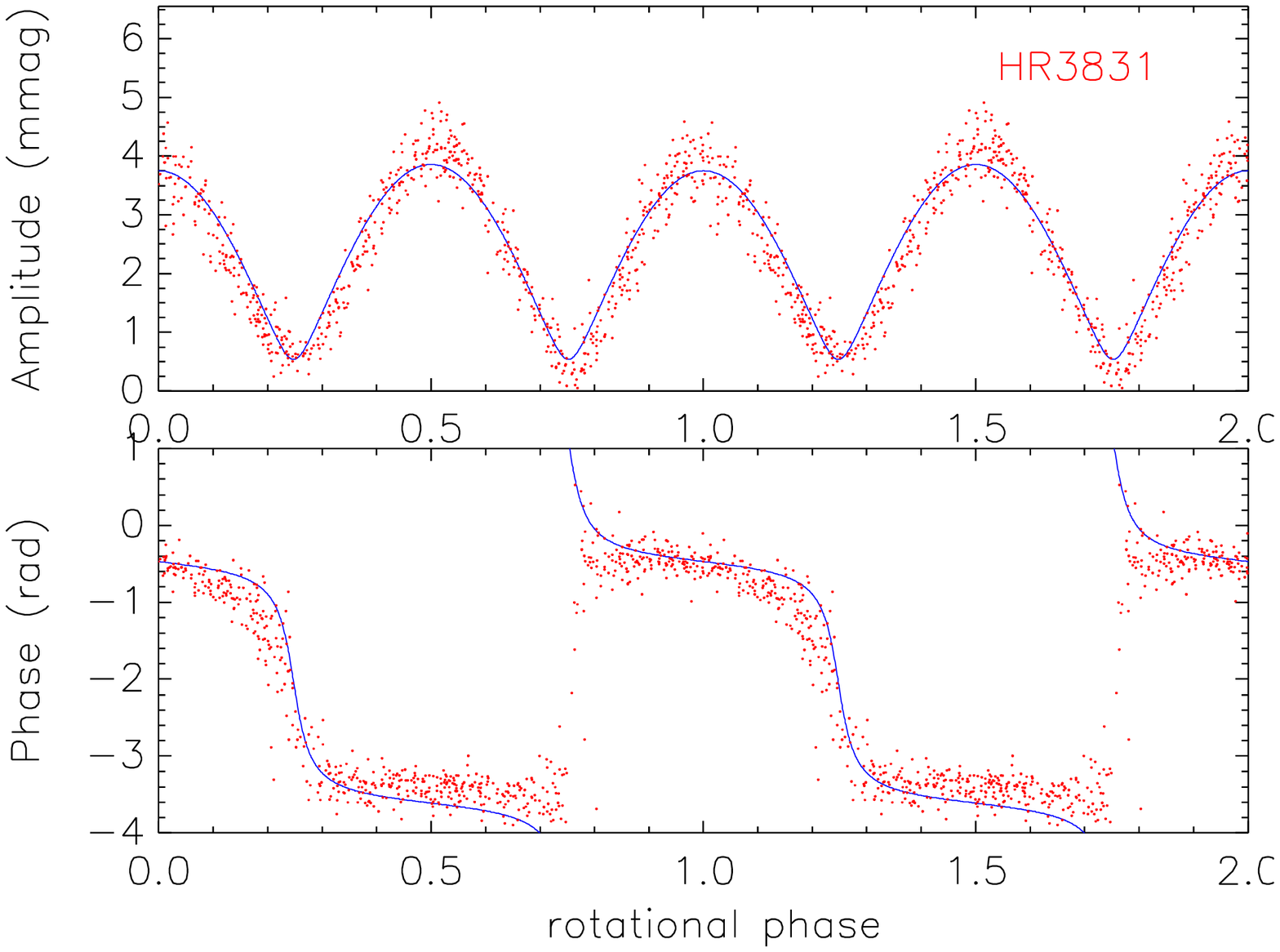}
\includegraphics[width=6.5cm,angle=0]{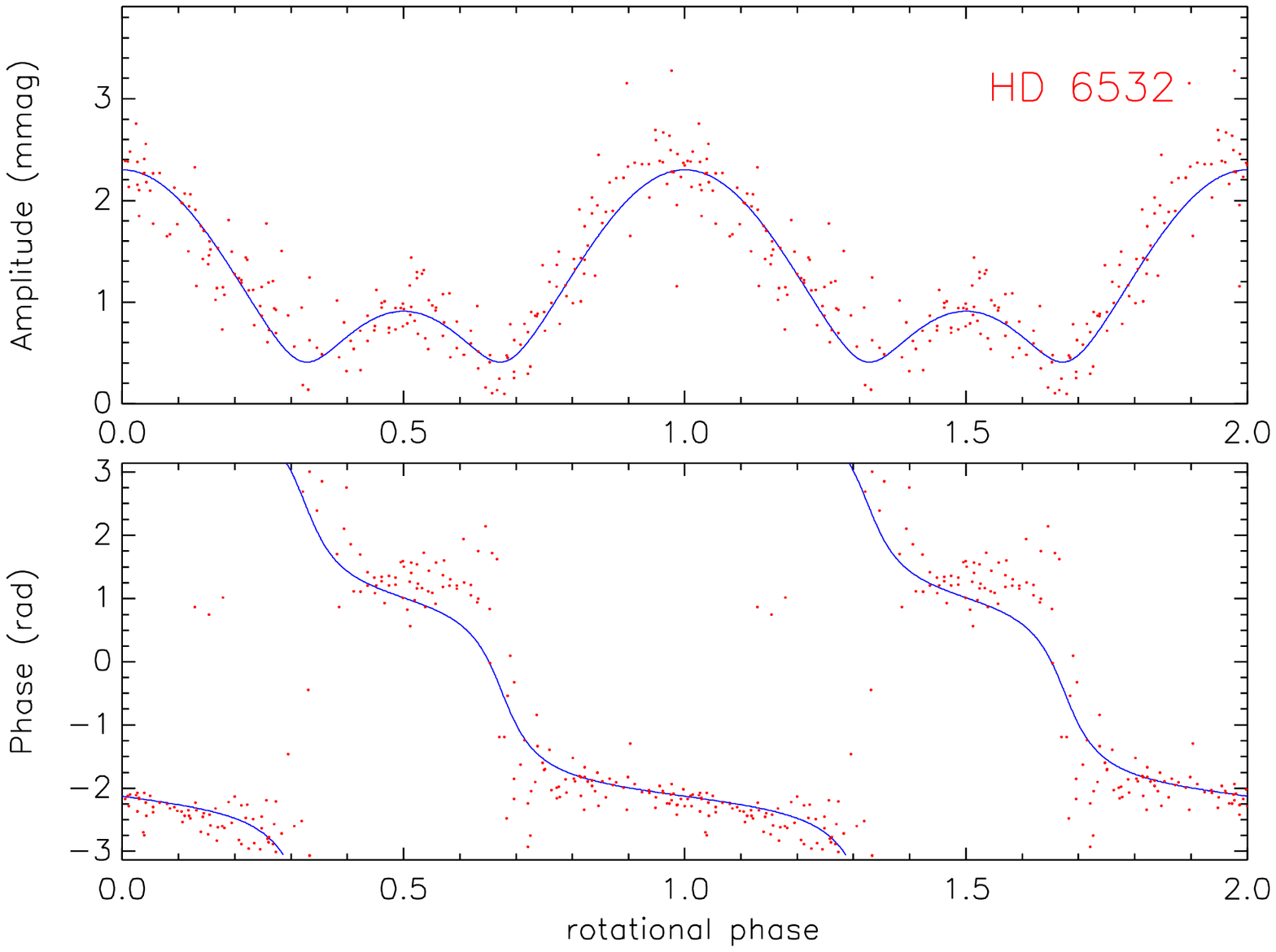}
\includegraphics[width=6.5cm,angle=0]{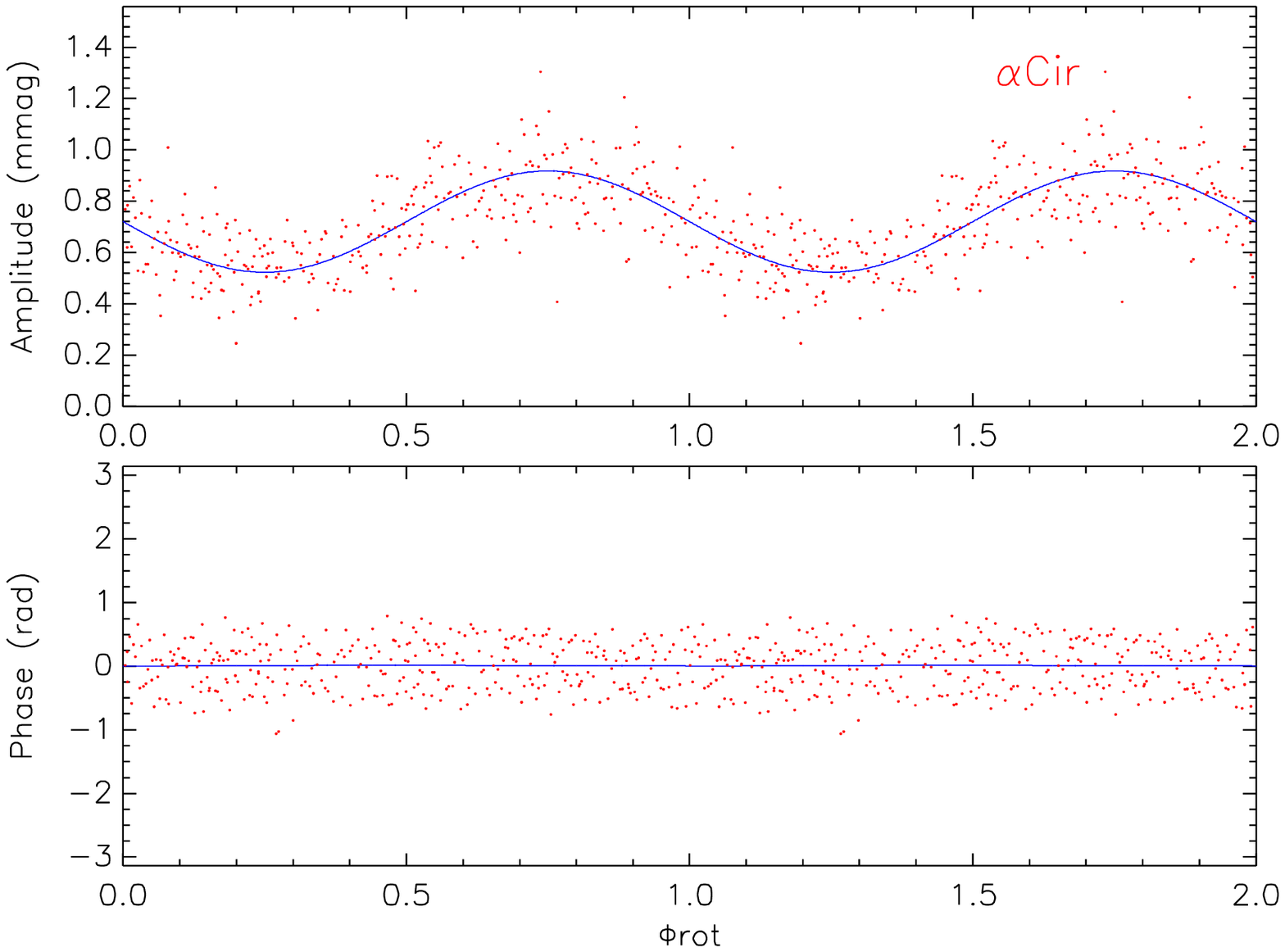}
\includegraphics[width=6.5cm,angle=0]{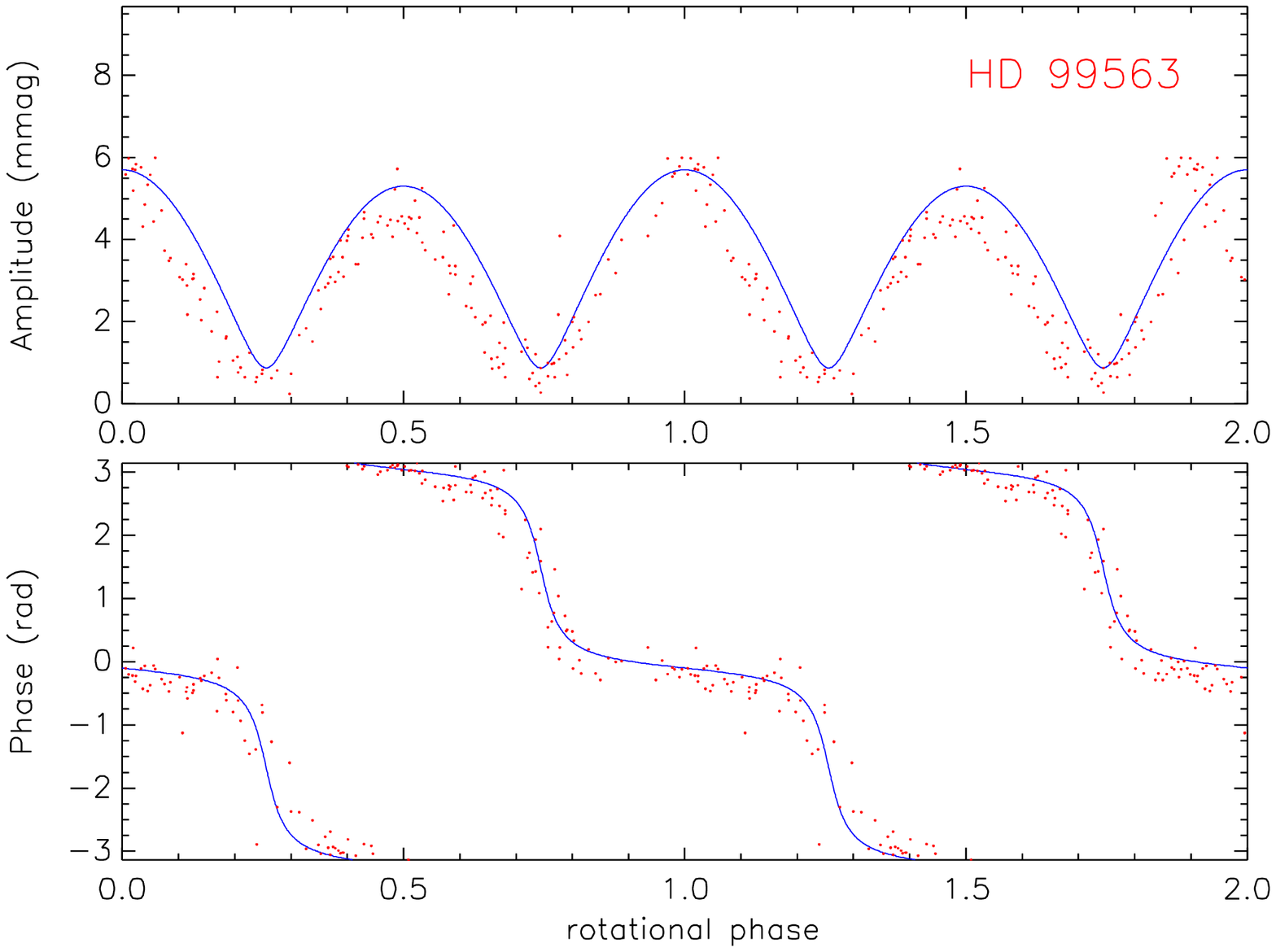}
\caption{Applications to four roAp stars: HR\,3831 (upper left), HD\,6532 (upper 
right), {$\alpha$}\,Cir (lower left), and HD\,99563 (lower right).  The dots 
represent the observed values and the full lines the theory. Observational data 
are from Kurtz et al. (1997a) for HR\,3831, Kurtz et al. (1996a) for HD\,6532, 
Bruntt et al. (2009) for {$\alpha$}\,Cir, and Handler et al. (2006) for HD\,99563. 
In our linear treatment of pulsation, the absolute amplitude of the mode is 
arbitrary. Therefore, the amplitudes of the light curves are scaled so that the 
observed ones equal the theoretical ones at $\Phi=0$. The theoretical 
pulsation phase is with respect to an arbitrary zero-point in time, so is shifted 
to reproduce the observed pulsation phase at rotation phase $\Phi=0$. The best-fits to these curves are ($i=68^\circ$, $\gamma = 91^\circ$, $\psi=-7^\circ$) for 
HR\,3831, ($i=40^\circ$, $\gamma = 70^\circ$, $\psi= -15^\circ$) for HD\,6532 
($i=37^\circ$, $\gamma = 20^\circ$, $\psi=0.5^\circ$) for $\alpha$\,Cir, and 
($i=44^\circ$, $\gamma = 86^\circ$, $\psi=- 9^\circ$) for HD\,99563. The different 
shapes of the amplitudes are caused by different inclinations of the mode axes 
with respect to the line-of-sight of the observer. The four modes are almost 
linearly polarized. The rotational variabilities of the magnetic fields of HD\,6532, 
$\alpha$\,Cir, and HD\,99563 are not known; such observations will thus test our 
model for agreement with the angles we have given.}
\label{fig:delta_real}
\end{center}
\end{figure*}

We apply these considerations to the roAp star HR\,3831 for which we have good 
time measurements for both magnetic and pulsation data. This star pulsates in a 
single mode with a frequency of $\omega/2\pi = 1427\,\mu$Hz (Kurtz et al. 1997a). 
From photometric data, the mode appears essentially as a dipole since there is a 
clear phase reversal of $\pi$ at the minima of the rotational envelope. 
Equivalently, this appears in the frequency domain as a dominant triplet with 
peaks that are exactly separated by the rotation frequency, and is clear evidence of a dipole mode 
pulsation. Extra $\ell \neq 1$ components are present in the spectrum, which 
indicates a distortion of the dipole by the magnetic field (Kurtz et al. 1997a), but they appear small in photometry, possibly because of either a slight influence  of the magnetic field or a geometrical cancellation (Saio \& Gautschy 2004 sec. 5.4) of the distorted mode.

The radial velocity measurements of Kochukhov (2006) indicate that  the Nd\,\textsc{iii} 
lines have a strong $\ell=3$ component, much larger than in photometry, which is in 
strong disagreement with the mode geometry assumption of $\ell=1$ made in section 2. We emphasize however that the 
Nd\,\textsc{iii} lines are formed very high in the atmosphere -- $\tau_{5000} 
\approx 10^{-4}-10^{-5}$ -- far above the continuum-forming region $\tau_{5000} 
\approx 1$ that we observe in photometry, hence the two sets of observations 
do not sample the same layers of the star. The Nd\,\textsc{iii} line formation is 
above the outer turning point of the mode reflection and therefore radial velocity 
measurements of this line do not correspond to the trapped magneto-acoustic modes 
but rather to the running part of mode in the atmosphere.

As mentioned in section 2, our formalism is applicable to the modes that are weakly distorted by magnetic field such that they can be represented by a single value of $\ell=1$. We therefore assume  that it is the case for HR 3831.
Our formalism enables us to 
determine the mode inclination with respect to the rotation axis by 
comparing with observed light curves. One needs an independent determination of the 
magnetic field axis orientation to conclude about the alignment of the mode with magnetic 
axis.

The magnetic field of HR\,3831 was first measured by Thompson (1981), who found a 
dipole field and determined the star's magnetic ephemeris. These magnetic data and a large 
set of pulsation data were analysed by Kurtz et al. (1992), who found that the 
time of the magnetic minimum coincides with the time of the pulsation maximum. 
More precisely, the weakest of the two maxima of the rotational envelope of 
pulsation coincides with the positive extremum of the magnetic field variations.

More recent magnetic observations were carried by Mathys (1995), Bagnulo et al. 
(1999) and Kochukhov et al. (2004). They found drastically different results 
concerning both the polar magnetic strength ($B_p$) and orientation toward the 
rotation axis ($\beta$). In the first two papers, the values of the field strength 
are considerably stronger ($B_p\sim 11$ kG) than that of Kochukhov et al. ($B_p \sim 
2.5$ kG). The orientations with respect to the rotation axis are also very 
different with $\beta = 7 \dg$ in Bagnulo et al. (1999) and $87\dg$ in Kochukhov 
et al. (2004). These differences are discussed in Kochukhov et al. (2004) who 
argued that these discrepancies are probably caused by the neglect of  Mathys and Bagnulo 
et al. of the presence of stellar spots in their analysis.

We emphasize at this stage that the magnetic measurements of Bagnulo et al. (1999) 
appear incompatible with the presence of dipole-like pulsation aligned with the 
magnetic axis. That would indeed imply an improbably high pulsation amplitude 
that is not believable: The observed amplitude of the pulsation would be 
$R(\lambda = 82^\circ) = R(\lambda = 0^\circ) \cos(82)$. The observed semi-
amplitude for HR\,3831 is about 5\,mmag in $B$; with this geometry, if the mode 
could be seen pole-on ($\lambda = 0$), then it would have an intrinsic semi-
amplitude of 36\,mmag! This is vastly larger than the largest amplitude seen for 
any of the known roAp star of 8\,mmag, i.e. in the case of HD\,60435, and that is for 
multiple modes beating with each other.

In their work, Bigot \& Dziembowski (2002) determined the magnetic configuration 
($B_p, \beta, i$) of the star by fitting the relative amplitudes of the triplet. 
With these parameters, they derived the mode inclination. However, this approach 
leads to two independent constraints, and  the authors had to rely on an independent determination 
of the inclination to the line of sight $i$. At that time, this value was 
available in Bagnulo et al. (1999), who proposed that $i=89^\circ$. Using that value of 
$i$, Bigot \& Dziembowski identified  a mode that is almost linearly polarized along the $X$-axis 
($\psi$ is small) but with an axis of symmetry that is steeply inclined  to the 
magnetic or rotation axes, $35^\circ $ and $42^\circ$, respectively. This result 
was surprising because it suggested for the time that modes in roAp may not be 
aligned with the magnetic field axis. This result was questioned by Kochukhov 
(2006), who found a mode nearly that is aligned with the magnetic axis from radial 
velocity measurements. He argued that the geometrical model of Bigot \& 
Dziembowski (2002) is too simple to account for the complexity of pulsation in 
these objects.

We however emphasize that the validity of the formalism in Bigot \& Dziembowski 
(2002) cannot be questioned by this argument, the two studies
not being based on the same observational constraints. A different value of $i$ leads to a completely different 
value of the mode inclination. As a matter of fact, if we assume the value proposed by Kochukhov (2006), i.e. $ 
i=68 \dg$, the geometrical model fits the light curve variations very well for 
both the amplitudes and phases, as seen in Fig.\,\ref{fig:delta_real}. The mode is 
now inclined by $\gamma = 91 \dg$ with respect to the rotation axis, i.e. 
almost aligned with the magnetic axis (if we assume the value of $\beta = 87 \dg$ 
given by Kochukhov et al. 2004). This mode is still found to be almost linearly 
polarized, i.e. with a small non-axisymmetric components, since $|\psi| \approx 7 
\dg$.  The ratio of magnetic to centrifugal shifts of frequencies is $\mu \approx -0.8$. As shown in Fig. 2 of Bigot \& Dziembowski (2002), 
even if magnetic and centrifugal shifts are comparable ($|\mu|\approx 1$), 
the modes can be almost linearly polarized along the magnetic axis if $\beta$ is large.
At the time of the positive magnetic extrema, the angle of the pulsation 
axis with the line of sight is $\alpha(0)=23\dg$ and half-rotation period later 
$\alpha(\pi)=159\dg$. These results are in good agreement with those presented in 
Kochukhov (2006).

Since the data of Bagnulo et al. (1999) are unlikely, we conclude that the 
pulsation mode in HR\,3831 is nearly aligned with the magnetic axis. 


\subsection{Application to HD\,6532}

HD\,6532 is one of the few roAp stars to show a polarity-reversing dipole-like 
pulsation mode (Kurtz et al. 1996a); the others are HR\,3831 discussed in the last 
section, HD\,99563 (Handler et al. 2006, and below), and HD\,80316 (Kurtz at al. 
1997b). The star HD\,6532 has a short rotation period of only 1.944973\,d (Kurtz et al. 
1996b), hence a relatively rapid rotational velocity, similar to that of HR\,3831. The 
geometry of the star is such that one pulsation pole is seen for about two-thirds of 
the rotation period, and the other for one-thirds, with a pulsation phase reversal at 
quadrature. With this geometry, one pole is seen at a far more favourable aspect 
than the other, hence at larger  amplitude.

Little is known about the magnetic field of HD\,6532. Mathys \& Hubrig (1997) did 
not detect the longitudinal field, but found a significant quadratic field of $22 
\pm 4$\,G. Magnetic measurements are not easy to make for HD\,6532, given its 
relative faintness ($V=8.4$) and rotationally-broadened lines. We are thus unable 
to compare our pulsation solution to the light curves of HD\,6532 with independent 
magnetic data, but we can make predictions about the geometry of the star that can 
be tested by future magnetic measurements.

We  modelled the pulsation data for HD\,6532 taken from Kurtz et al. (1996a). The 
best fit for we have obtained is ($i=40^\circ$, $\gamma = 70^\circ$, 
$\psi=-15^\circ$). The mode axis is closer to the line-of-sight than in HR\,3831, 
i.e. $\alpha(0)= 30^\circ$ and $\alpha(\pi) = 110^\circ$. The difference between 
the two maxima is therefore larger for this star than for HR\,3831; we found that
$R(0)/R(\pi) \approx  2.5$, as seen in Fig. \ref{fig:delta_real}. With these values of 
$\alpha$, the observer sees the two hemispheres as the star rotates, as in 
HR\,3831. The important point here is that we can match the rotational amplitude 
and phase variations seen in the light curves of HD\,6532 with our model and make 
clear predictions about the geometry that can be tested against future magnetic 
measurements.

\subsection{Application to $\alpha$\,Cir}

The star $\alpha$\,Cir is the brightest of the roAp stars ($V=3.2$) and has basically 
a single pulsation mode with a frequency of 2.442\,mHz ($P = 6.8$\,min) (Kurtz et al. 
1994; Bruntt et al. 2009). Several other modes of consecutive odd-even degrees are 
present but with smaller amplitudes (Bruntt et al. 2009). These authors derived a 
larger value of the large separation $\Delta \nu=60\,\mu$Hz than the previous 
estimate (Kurtz et al. 1994). This new value of $\Delta\nu$, combined with a drastic 
downward revision of the effective temperature (7400\,K) thanks to an 
interferometric radius determination (Bruntt et al. 2008), leads for the first 
time to consistent asteroseismic and Hipparcos parallaxes, in contrast to the 
result found by Matthews et al. (1999). There had been a long history of attempts to 
find the rotation period of $\alpha$\,Cir until it was finally determined by an analysis of 
the pulsation data; the rotation period is $4.4790 \pm 0.0001$\,d (Bruntt et al. 
2009). The rotational sidelobes in the amplitude spectrum for this star are small, 
indicating that there is little modulation of the amplitude, hence this aspect of 
the pulsation mode does not vary much  with rotation.

That is also true for the magnetic field, for which no rotational variations have therefore been 
detected. Mathys \& Hubrig (1997) detected no longitudinal field, but found 
a quadratic field strength of 7.5\,kG. As for HD\,6532 above, we cannot test our 
model against independent magnetic measurements, but the geometry that we find can be 
tested against future, higher precision magnetic measurements.

We  modelled pulsation data for $\alpha$\,Cir taken from Bruntt et al. (2009); 
these are white light data taken by the WIRE satellite that have a higher 
precision than the $B$ data of Kurtz et al. (1994), but otherwise have the same 
rotational modulation. We used the value of $i=37\dg$ found by Bruntt et al. 
(2008) by combining an interferometric determination of the radius with the 
determination of $v \sin i$ by line spectroscopy. Our best-fit is ($\gamma = 
20^\circ$, $\psi=0.5^\circ$). In this case, the mode axis is still almost linearly 
polarized along the rotation axis ($\gamma=20^\circ$). The angles made by the 
pulsation axis to the line-of-sight, at the time of the two maxima, are close, 
i.e. $\alpha(0)=17\dg$ and $\alpha(\pi)=57\dg$. Owing to the small tilt of the 
pulsation axis with respect to the rotation one, the observer always sees the same 
hemisphere and there is no phase reversal. The phase is therefore nearly constant. 
The ellipticity $\psi$ is very small for this star with a value of $0.5^\circ$. 
This mode is very close to an $m_R=0$ mode.

\subsection{Application to HD\,99563}

The star HD\,99563 is a singly-periodic roAp star pulsating in a distorted dipole mode 
that shows both pulsation poles over the rotational cycle. Photometrically, it is 
one of the larger amplitude roAp stars with a semi-amplitude as high as 6\,mmag in 
Johnson $B$ (Handler et al. 2006). Its radial velocity variations have the highest 
amplitudes seen for an roAp star, reaching 5\,km\,s$^{-1}$ for some 
Eu\,\textsc{ii} lines, as a consequence of their strong concentration in small 
abundance spots near the pulsation poles (Freyhammer et al. 2009), their 
stratification high in the atmosphere, and the intrinsically high pulsation 
amplitude of HD\,99563 (for an roAp star). The rotation period for HD\,99563 was 
determined by Handler et al. (2006) to be $P_{\rm rot} = 2.91179 \pm 0.00007$\,d. Some variation in both the amplitude and phase with rotation, plus  the 
strong $\ell = 3$ 
components found by Handler et al. show this star to have a magnetically distorted 
dipole mode. Freyhammer et al. (2009) showed that some rare earth elements are 
overabundant by more than $10^6$ solar values in spots, and that even H shows 
a nonuniform surface distribution as a consequence of the rare-earth element 
patches. The result of this is that the observations -- even in photometry -- do 
not sample the surface uniformly. The spatial filter caused by the extreme 
abundance anomalies is probably the source of the apparent distortion of the mode 
determined from photometry.

We modeled the multisite photometric observations of Handler et al. (2006) 
using the rotational inclination $i = 44^\circ$ determined by those authors. From 
this, we found that $\gamma = 86^\circ$ and $\psi=- 9^\circ$, as can be seen in Fig.\,4. 
This is the poorest fit of our model to the four stars tested, which we believe is 
a result of the strong distortion of the observations by the surface spatial 
filter. An interesting test of this would be to study the pulsation of HD\,99563 
in Johnson $V$, where there is little rotational variation in the mean light (see 
Fig.\,4 of Handler et al. 2006). While the pulsation amplitude will be smaller in 
Johnson $V$ than $B$, the star is of sufficiently high amplitude to ensure that data of good signal-
to-noise ratio can be obtained. We expect that such observations will show a weaker 
distortion. Magnetic measurements of HD\,99563 are consistent with the 
pulsation geometry (see Freyhammer et al. 2009 for a discussion).
As for HR\,3831, the mode axis is nearly aligned with the magnetic one ($\beta=88\dg$, see Freyhammer et al. 2009).

\subsection{Comparison  of the four roAp stars}

On the basis of this small sample, it appears that there is no preferential inclination 
of the pulsation axis to the rotation one since $\gamma = 91^\circ$, $70^\circ$, 
$20^\circ$, and $86^\circ$. The lack of magnetic data prevent any general 
conclusion regarding the alignment of the mode and magnetic axes. However, for the two roAp stars (HR\,3831, HD\,99563) for which we have the value of magnetic obliquity $\beta$, we found from the light curve
analysis that the modes are nearly aligned with the magnetic axis ($\gamma\approx \beta$). The different 
behaviour of the light curves is explained by the different inclinations of their 
axis of symmetry  to the line-of-sight. We note, however, that there 
is a tendency for the modes to have small ellipticities of  $|\psi| = 7^\circ$, 
$15^\circ$, $0.5^\circ$, and $9^\circ$. A new polarity-reversing roAp star 
found by the {\it Kepler} mission, KIC\,101095926 shows a similar small 
ellipticity of $|\psi| = 4^\circ$ (Kurtz et al., 2011). This means that most 
of the displacement vectors of these modes remain very close to the magnetic plane 
during the pulsation cycle. This may be a common property of roAp stars that 
should be investigated  by considering a much larger sample than the 
present one. 

\section{Conclusions}

We have presented  a formalism to explain the aspect of the light curves 
associated with dipole oscillations in roAp stars. In this purely geometrical 
model, all the complexity is reduced to three aspect angles, namely  the mode and 
observer inclinations with respect to the rotation axis and the polarization angle of 
the mode. This geometrical model can reproduce global mode properties and is 
particularly useful for asteroseismic diagnostics. As we have shown, the model works well 
with the four roAp stars considered in this paper. Owing to its simplicity, it has 
limitations and cannot reproduce detailed line profile variation or derived 
accurate field strength. This would need a more realistic rotation-magneto-
hydrodynamic numerical model.

We have shown that the light curve variations in roAp stars are strongly 
influenced by the magnetic field strength, which is parameterized in this study by 
a single parameter. The reason is that an increase in the field strength pushes 
the mode to be either aligned or perpendicular to the magnetic axis. This consequently 
changes the aspect of the projection of the mode onto the line-of-sight of the 
observer. We have also clearly shown that the coincidence between the observed 
pulsation and magnetic maxima is not evidence of an alignment of their respective 
axes as had been commonly thought.

We have applied this formalism to four roAp stars and in particular the  well-studied roAp star HR\,3831. For this star, we revised the  
conclusion of Bigot \& Dziembowski (2002), who derived a highly inclined dipole 
mode with respect to the magnetic axis. We do not question their formalism, but 
rather the magnetic data they used. Indeed, in either the light curve or triplet 
analysis, one needs to assume an inclination for the observer. The one used by 
Bigot \& Dziembowski (2002) was taken from Bagnulo et al. (1999) and led to well 
inclined mode. Using the more recent and the more reliable data of Kochukhov 
(2006), we have found a mode that is nearly aligned with the magnetic axis, quite close 
to the original picture of the oblique pulsator model of Kurtz (1982). This means 
that the magnetic field strength is strong ($\geq 1$ kG). In this respect, the 
value of 2.5 kG proposed by Kochukhov et al. (2004)  agrees with the 
present conclusions. This strength makes the centrifugal force inefficient to tilt 
significantly the mode axis far from the magnetic axis.

The geometrical picture of pulsation presented in Bigot \& Dziembowski (2002) and 
in this paper clearly explains the photometric data and leads to mode properties that 
are in good agreement with radial velocity measurements, at least in the case of 
HR\,3831. In this model, the inclination $\gamma$ and polarization $\psi$ are two 
{\it global} properties of the standing mode and therefore do not vary with depth. 
This is the converse of the result that Kochukhov (2006) found for radial velocity 
measurements of Pr and Nd lines. He claimed that the geometrical approach is 
inadequate since it does not account for this variation with depth. We, however, 
emphasize that the variation in the ratio of the peak amplitudes of the triplet  
found by Kochukhov (equivalently $\gamma$ and $\psi$) indicates the presence of a 
running wave travelling in the atmosphere far above the continuum-forming region 
where the fluctuations of luminosity originate. As shown by Saio et al. 
(2010) with a realistic magneto-hydrodynamic  pulsation model, the phase of the 
standing mode is constant in the deep atmosphere and in the interior, whereas the 
running wave in the atmosphere has a changing phase with depth. Our formalism 
applies to the global standing mode whose phase is therefore constant, and the 
parameters $\gamma$ and $\psi$ can be regarded as global geometrical parameters.

\begin{acknowledgements}
L. B. thanks W. Dziembowski for fruitful discussions at an early stage of this work and the referee, H. Saio, for his constructive suggestions.
\end{acknowledgements}

\appendix

\section{Special types of light curves}

We apply the general formalism developed in this paper to special 
values of the polarization given by
$\psi=0$, $|\psi|=\pi/2$, and $|\psi|=\pi/4$.

\subsubsection{Linear polarization $\psi=0$}

We consider the important case of modes linearly polarized in the plane 
(${\bf B,\Omega}$),  characterized by a value of $\psi=0$. This case, which is even 
idealistic, is not far from reality since it is believed that most of the 
pulsation modes in roAp stars are almost linearly polarized.

These dipole modes ($\psi=0$) always show extrema in phase with the magnetic 
extrema because their pulsation axes lie in the plane (${\bf B,\Omega}$). Their 
amplitude is written

\begin{equation}\label{eq:linearly_polarized}
R(\Phi)\propto\left |\cos\alpha (\Phi)\right |.
\end{equation}

To precisely examine the coincidence between magnetic and pulsation 
maxima, we have to consider four angles which are  the angles between the mode 
axis and the line-of-sight at the two extrema, $\alpha(\Phi=0)$ and 
$\alpha(\Phi=\pi)$, and the angle between the magnetic axis and the line-of-sight 
at the same phases, $\lambda (\Phi=0)$ and $\lambda (\Phi=\pi)$.

The conditions required to have maxima at $\Phi=0\,[\pi]$, i.e. Eq.\,\ref{eq:max0}-
\ref{eq:maxpi}, can be divided into the following three cases:
\begin{itemize}

\item Case I.: the mode shows two maxima at $\Phi = 0$ and $\Phi=\pi$, with phases 
$\Psi$ of opposite signs:
\begin{equation}\label{eq:condition}
|\alpha (0)|\leq\frac{\pi}{2}\leq |\alpha (\pi)|
\end{equation}
or
\begin{equation}
|\alpha (\pi)|\leq\frac{\pi}{2}\leq |\alpha (0)|.
\end{equation}
In this  case, the observer sees the two hemispheres of the dipole as the star 
rotates. The light curves resemble those shown in Fig.\,\ref{fig:dipole}.

\item Case II.: the mode shows one maximum at $\Phi=0$ and a minimum at 
$\Phi=\pi$, with the same signs for phases $\Psi$ :
\begin{equation}\label{eq:twomax}
|\alpha (0)|\leq |\alpha(\pi)|\leq\frac{\pi}{2}
\end{equation}
or
\begin{equation}
\frac{\pi}{2}\leq |\alpha (0)|\leq |\alpha(\pi)|.
\end{equation}

\item Case III.: the mode shows one maximum at $\Phi=\pi$ and a minimum at 
$\Phi=0$, with the same signs for the phases $\Psi$ :

\begin{equation}
|\alpha (\pi)|\leq |\alpha(0)|\leq\frac{\pi}{2}
\end{equation}

or

\begin{equation}
\frac{\pi}{2}\leq |\alpha (\pi)|\leq |\alpha(0)|.
\end{equation}

\end{itemize}

\noindent Cases II and III correspond to orientations of the system where the observer 
sees only one hemisphere as the star rotates. Examples of cases I and III are 
illustrated in Fig.\,\ref{fig:dipole}.

We also note  that for some specific inclinations of either the rotation axis or the mode 
axis, the light curve shows some extrema of the same amplitude. In the case of $i 
= 0 [\pi/4]$ and/or $\gamma = 0 [\pi/2]$, we have $|\alpha(0)| = |\alpha (\pi)|$, 
and thereby $R(0) = R(\pi)$.

The coincidence between the magnetic and pulsation maxima can occur in many cases 
depending on the mode and the magnetic field orientation. For example, the 
condition to have pulsation maxima at $\Phi=0$ and at $\Phi=\pi$ and a magnetic 
positive maximum at $\Phi=0$ together with a negative magnetic minimum at 
$\Phi=\pi$ combines the relations Eq.\,\ref{eq:condition} and Eq.\,\ref{eq:mag1}, 
which can be written

\begin{equation}\label{eq:rela}
|i-\gamma |\leq\frac{\pi}{2}\leq |i+\gamma|\hspace{0.6cm}{\rm and}\hspace{0.6cm} 
|i-\beta|\leq \frac{\pi}{2} \leq |i+\beta|.
\end{equation}

\noindent This relation is not very restrictive and allows a wide range of values 
of $(i,\gamma,\beta)$.

\begin{figure*}[th!]
\begin{center}
\vbox{\hbox{\includegraphics[width=7cm,angle=0]{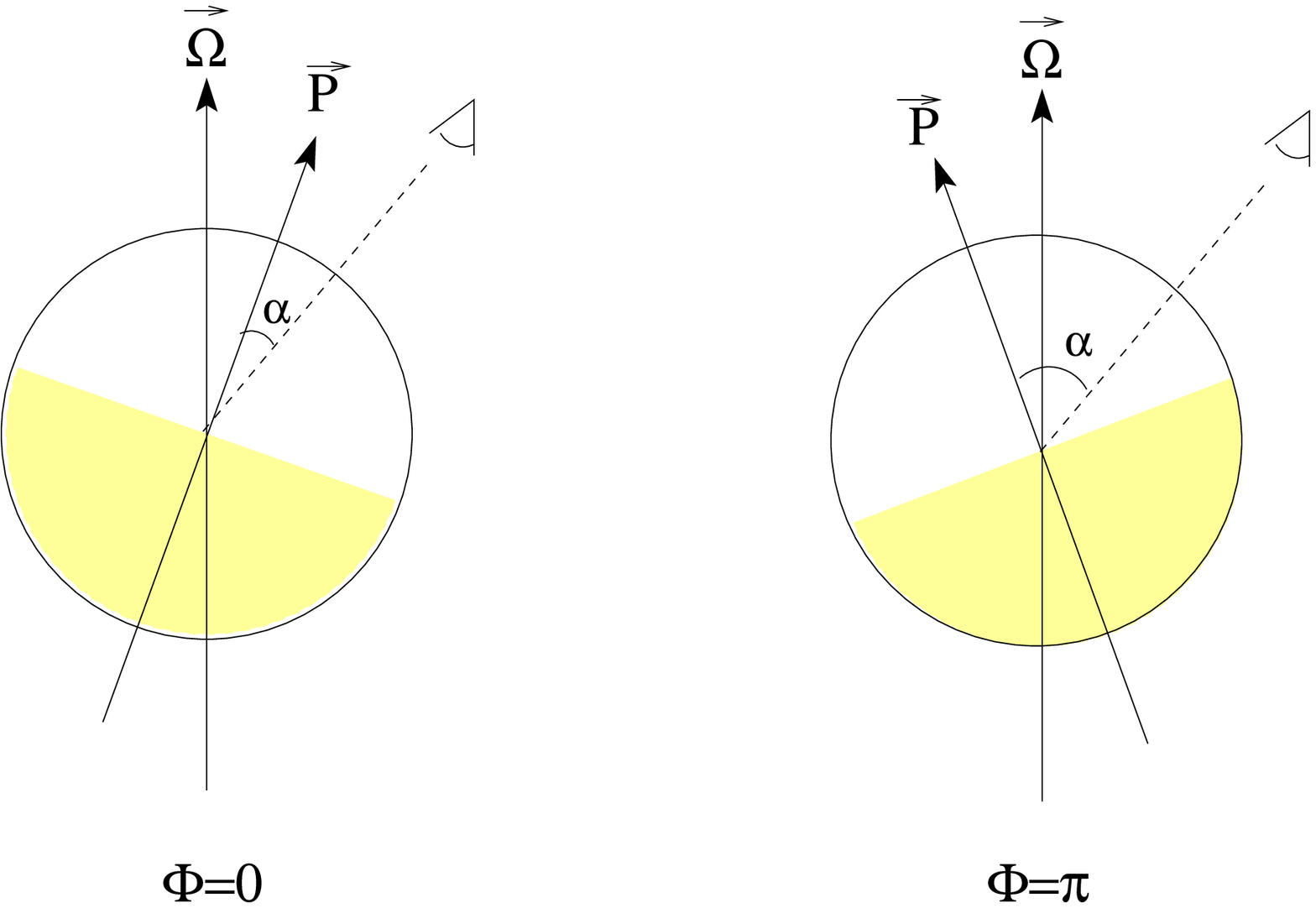} 
\hspace{2.0cm}\includegraphics[width=7cm,angle=0]{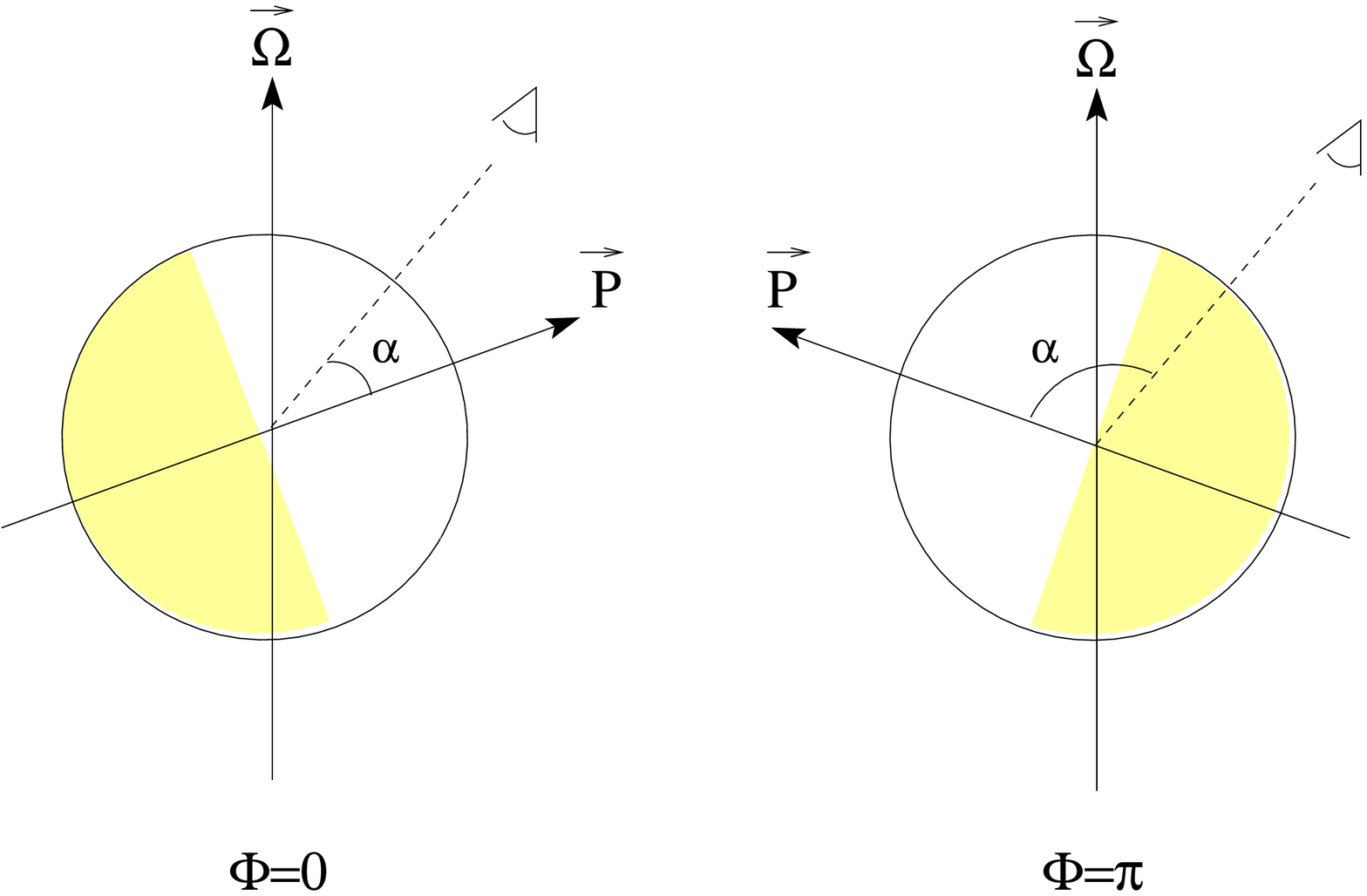}}
\vspace{0.5cm}\hbox{\includegraphics[width=7cm,angle=0]{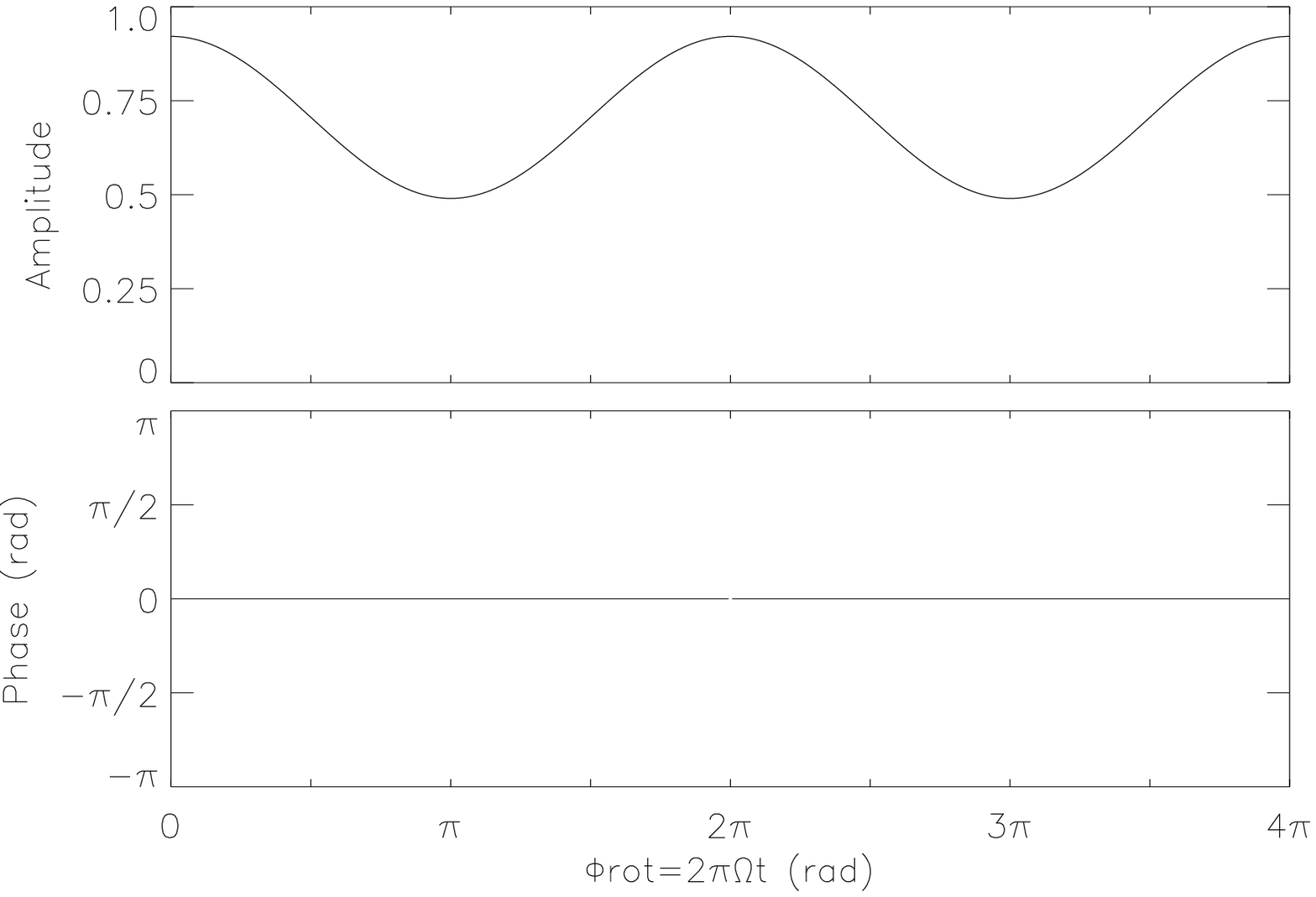} 
\hspace{2.0cm}\includegraphics[width=7cm,angle=0]{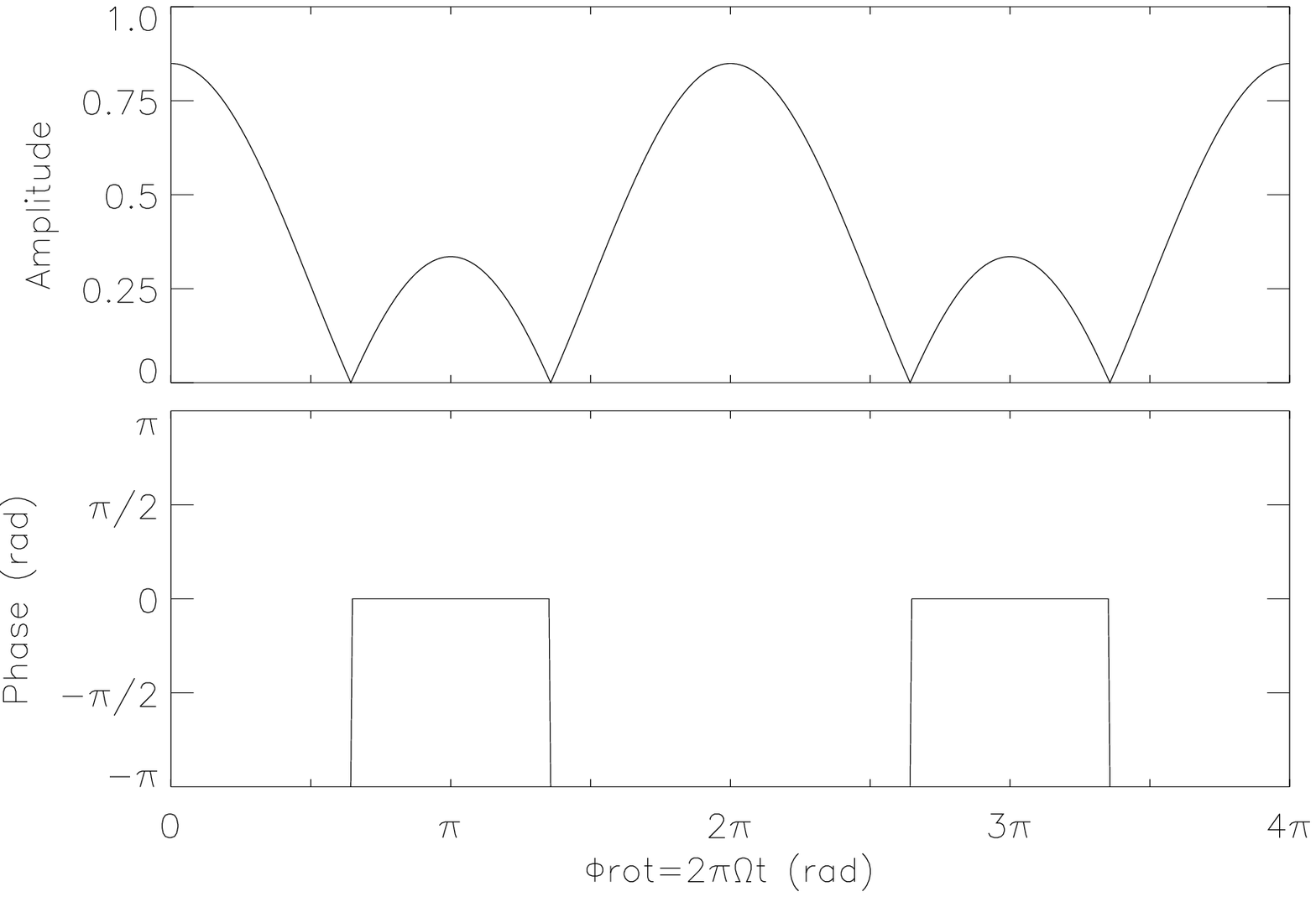}}}
\caption{Representation of the light curves for linear polarized ($\psi=0$) dipole 
modes with different inclinations of the pulsation axis. This illustrates the 
cases  discussed in section A.0.1. In both figures, the position of the 
observer is fixed, $i=40^o$. Left panel: the dipole axis remains close to 
the line-of-sight during the rotation of the star. The observer only sees one 
hemisphere of the mode, and thereby does not observe a phase jump during the 
rotation period. Hence, a maximum and minimum of the rotational envelope of the 
pulsation are observed at $\Phi=0$ and $\Phi=\pi$. Right panel: the 
dipole axis is now sufficiently inclined with respect to the line-of-sight so that 
the two hemispheres of the dipole can been seen by the observer. A phase jump by 
$\pi$ radians occurs when the node of the oscillations crosses the plane (${\bf 
i,\Omega}$). In this case, the amplitude of the rotational envelope leads to two 
maxima of different amplitudes ($\alpha(0) <\alpha(\pi)$).}
\label{fig:dipole}
\end{center}
\end{figure*}

\subsubsection{Linear polarization $|\psi|=\pi/2$}

In the case of linearly polarized modes orthogonal to the plane $({\bf 
B,\Omega})$, i.e. with $|\psi|=\pi/2$, we have

\begin{equation}\label{eq:linearly_polarizedbis}
R(\Phi)\propto |\sin\Phi|.
\end{equation}

\noindent No possible coincidence between magnetic and pulsation maxima can occur 
because the mode pulsates always in a direction orthogonal to the magnetic axis. The 
magnetic and pulsation maxima are always in quadrature.

\subsubsection{Circular polarization $|\psi| =\pi/4$}

In the case of $m=\pm 1$ modes in the rotation reference system, ($\psi=\pm\pi/4$, 
$\gamma =\pi/2$), the light curve simplifies to the simple expression

\begin{equation}\label{eq:linearly_polarizedbis}
R(\Phi) = |\sin i|\hspace{0.5cm}\Psi (\Phi) =\pm\Phi.
\end{equation}

\noindent The phase of the light curve then takes the characteristic shape of 
straight lines of slope $\pm 1$. This situation appears in Fig.\,\ref{fig:delta}.

\section{Nature of extrema at $\Phi=\Phi_o$}

The condition to ensure that the maxima are located at $\Phi=\Phi_o$ is written

\begin{equation}
(\tau^2-1)\cos 2\Phi_o <\xi.
\end{equation}

\noindent The maxima correspond to the instants when the $y$-axis of the 
ellipse is the closest to the plane $({\bf i,\Omega })$. We emphasize here that 
because of the orientation of the system (rotation axis and mode), these extrema 
are not located exactly at $\Phi=\pi/2\,[\pi]$, as we might think intuitively, but 
rather are phase shifted, except for $\gamma =\pi/2$ or $i =\pi/2$ for which 
they are located at $\Phi=\pi/2\,[\pi]$. Their amplitudes relative to the extrema 
at $\Phi=0$ are

\begin{equation}
\frac{R(\Phi_o^{\pm})}{R(0)} =\left |\frac{\tau}{\xi+1}\right 
|\sqrt{\frac{\xi^2}{\tau^2-1}+\tau^2},
\end{equation}

\noindent which do not vanish as long as the mode is not linearly polarized in the 
plane $({\bf B,\Omega})$, i.e. $\tau=0$.

\end{document}